\documentclass[twocolumn]{article}
\usepackage[utf8]{inputenc}

\usepackage{geometry}
\usepackage[dvipsnames]{xcolor}
\usepackage{soul}
\usepackage{subcaption}
\usepackage{booktabs}
\usepackage{adjustbox}
\usepackage{multirow}
\usepackage[all]{nowidow}
\usepackage{float}

\usepackage{amsmath}
\usepackage{amssymb}
\usepackage{bm}
\usepackage{nicefrac}
\usepackage{mathtools}

\usepackage[separate-uncertainty=true,parse-numbers=true]{siunitx}

\usepackage{pgfplots}
\usepackage{tikz}
\pgfplotsset{compat = newest}
\usetikzlibrary{quotes,angles,calc,arrows,patterns,quotes,external}
\usepgfplotslibrary{groupplots}
\usepackage[colorlinks=true, citecolor=blue,urlcolor=blue,linkcolor=blue,filecolor=black]{hyperref}
\usepackage[capitalise,nameinlink,noabbrev]{cleveref} % after amsmath

\usepackage{authblk}

\usepackage{algorithm}
\usepackage{algpseudocode}

\usepackage[T1]{fontenc}

\title{High Purcell-enhancement in quantum-dot hybrid circular Bragg grating cavities for GHz-clockrate generation of indistinguishable photons}
\author[1,+]{Lucas Rickert}
\author[1,+]{Daniel A. Vajner}
\author[1,+]{Martin von Helversen}
\author[1]{Johannes Schall}
\author[1]{Sven Rodt}
\author[1]{Stephan Reitzenstein}
\author[2,3]{Hanqing Liu}
\author[2,3]{Shulun Li}
\author[2,3]{Haiqiao Ni}
\author[2,3,*]{Zhichuan Niu}
\author[1,*]{Tobias Heindel}
\affil[+]{These authors contributed equally to this work}
\affil[1]{Institut für Festkörperphysik, Technische Universität Berlin, Hardenbergstraße 36, 10623 Berlin, Germany}
\affil[2]{Key Laboratory of Optoelectronic Materials and Devices, Institute of Semiconductors, Chinese Academy of Sciences, Beijing 100083, China}
\affil[3]{Center of Materials Science and Optoelectronics Engineering, University of Chinese Academy of Sciences, Beijing 100049, China}
\affil[*]{Corresponding authors: tobias.heindel@tu-berlin.de, zcniu@semi.ac.cn}
\date{}

\begin{document}

%\maketitle

\twocolumn[
\begin{@twocolumnfalse}
\maketitle
\begin{abstract}
\noindent
We present Purcell-enhanced ($F_\mathrm{P}>25$) semiconductor InAs quantum dot decay times of $T_1<30$~ps, enabled by deterministic hybrid circular Bragg gratings (hCBGs). We investigate the benefits of these short $T_1$ times on the two-photon indistinguishability for quasi-resonant and strictly resonant excitation, and observe visibilities $\geq96\%$ at 12.5~ns time delay of consecutively emitted photons. The strongly Purcell-enhanced decay times enable a high degree of indistinguishability for elevated temperatures of up to 30~K, and moreover, allow for excitation of up to 1.28~GHz repetition rate. Our work highlights the prospects of high Purcell-enhanced solid-state quantum emitters for applications in quantum information and technologies operating at GHz clock-rates. 
\end{abstract}
\hspace{2cm}
\end{@twocolumnfalse}
]

\section{Introduction}
\label{sec:Intro}

Indistinguishable photons are a key component for implementing advanced photonic quantum information protocols. Higher-order entangled state generation enables boson sampling~\cite{aaronson_computational_2011} while remote-party multi-photon interference is a prerequisite for implementation-relevant quantum key distribution (QKD) protocols, like measurement- and device-independent QKD~\cite{zapatero_advances_2023-1}. First proof-of-concept demonstrations of device-independent QKD protocols based on indistinguishable photons from trapped atoms have been reported~\cite{zhang_device-independent_2022}, but a significantly higher degree of scalability is expected by realizing reliable on-demand generation of indistinguishable single photons from solid-state two-level-systems~\cite{rodt_deterministically_2020}.

While there have been several quantum emitters~\cite{aharonovich_solid-state_2016} enabling the emission of sub-Poissonian light, so-called solid-state artificial atoms formed by semiconductor quantum dots (QDs)~\cite{vajner_quantum_2022}, remain the only platform that simultaneously combines low multi-photon probabilities~\cite{schweickert_-demand_2018}, emission of entangled photon pairs~\cite{benson_regulated_2000,akopian_entangled_2006}, near-ideal brightness and high degrees of indistinguishability of consecutively emitted photons~\cite{ding_high-efficiency_2023,tomm_bright_2021,somaschi_near-optimal_2016}. This makes them ideal candidates for applications in quantum information technologies~\cite{heindel_quantum_2023}. Especially the emission of highly indistinguishable photons is a challenge, considering that QDs are hosted in solid-state environments, which can cause significant decoherence for the generation of emitted photons due to the influences by phonons~\cite{bylander_interference_2003,denning_phonon_2020} in the host matrix and also spectral diffusion due to charge carriers in the emitter's vicinity~\cite{liu_single_2018}, often indicated by blinking.

Typical mitigation strategies are cooling to helium-cryogenic temperatures and external charge control~\cite{somaschi_near-optimal_2016,zhai_low-noise_2020}, as well as coherent excitation mechanisms~\cite{muller_resonance_2007,ates_post-selected_2009} to ensure the emission of coherent, indistinguishable photons. With these measures, remote two-photon interference has recently been demonstrated from QDs at record-high visibilities of 93\%~\cite{zhai_quantum_2022}.

Another possibility to minimize the influence of the solid-state environment on the indistinguishability is to decrease the decay time $T_\mathrm{1}$ of the single-photon generating transition, which can be realized by integrating the QD into a photonic cavity and making use of the Purcell effect~\cite{purcell_e_m_b10_1946}. Reduced $T_1$-times increase the two-photon indistinguishability in the presence of inhomogeneous broadening~\cite{gold_two-photon_2014}, including dephasing from phonons~\cite{grange_cavity-funneled_2015,grange_reducing_2017}, potentially enabling operation without cost-intensive liquid-helium infrastructure.  

Short $T_1$ times are furthermore relevant when aiming to excite the QD at high excitation rates. Here, $T_1$ should be ideally considerably shorter than the inverse repetition rate $1/f$ (for example, $1/f\geq 5\cdot{T_1}$), to avoid overlapping of subsequent pulses from following excitation circles, which would require temporal filtering leading to a reduced efficiency due to the discarding of photon events or an increased multi-photon probability. 
In the past, mostly electrical excitation due to fast possible excitation rates were employed for GHz-driving of In(Ga)As-QD systems, but this limited indistinguishability and a significant amount of multi-photon pulses remained, due to the non-resonant excitation condition and considerable pulse overlap~\cite{hargart_electrically_2013,schlehahn_electrically_2016,shooter_1ghz_2020}. Optically excited InAs/InP QDs have been reported at GHz clock-rates, but with the typical $T_1$-times, multiphoton-suppression at $f\geq1$\,GHz was also limited here due to overlap of consecutive pulses~\cite{anderson_gigahertz-clocked_2020}. 
These limitations have recently been overcome in a work using coherent optical GHz excitation to generate entangled photons from a GaAs-QD~\cite{hopfmann_maximally_2021}. However, no Purcell enhancement was employed here, but rather the faster $T_1$-time of this material system. Coherent GHz driving of InGaAs QDs with their inherently longer $T_1$-times has been elusive so far.   

To enable Purcell-enhanced $T_1$-times suited for GHz-operation, close-to-ideal overlapp between the quantum emitter and the cavity mode must be achieved both in the spatial and spectral domain, a challenge hardly mastered to date using non-deterministically fabricated QD devices and resulting in low yield~\cite{wang_-demand_2019}. Photonic cavity systems like micro-pillars~\cite{somaschi_near-optimal_2016,ding_-demand_2016}, photonic crystal cavities~\cite{liu_high_2018} or (hybrid) circular Bragg grating cavities (hCBGs)~\cite{yao_design_2018,rickert_optimized_2019,liu_solid-state_2019} benefit from or even require typically spatial emitter integration precision of well below 50$\,$nm for significant Purcell enhancement.

The latter hCBG cavities have gained considerable research interest recently, since they allow for simultaneous out-coupling- and Purcell-enhancement of multiple transitions of the same QD emitter due to their low-Q and broadband cavity properties~\cite{wang_-demand_2019,liu_solid-state_2019}. They furthermore allow for post-growth optimization of the cavity properties by design modifications~\cite{rickert_optimized_2019}, which is only possible to a limited extent with e.g micropillar cavities.
While several QD-hCBG cavity integration methods are reported as deterministic, this often only implies that a QD is present in the fabricated cavity, but nonetheless significant deviation from theoretically possible Purcell factors is observed, generally attributed to a limited spatial integration precision. Several works have shown experimentally observed Purcell factors $F_\mathrm{P}$ of 3-7~\cite{liu_solid-state_2019,kolatschek_bright_2021,nawrath_bright_2023,holewa_high-throughput_2024,barbiero_high-performance_2022,kaupp_purcellenhanced_2023-1} with theoretically expected values of $F_\mathrm{P}\sim$20 from simulations. The highest reported measured Purcell enhancement for QD-hCBG cavities so far are 11.7~\cite{rota_source_2024} and 11.3~\cite{wang_-demand_2019}, respectively, corresponding to $T_1$-lifetimes of $14\,$ps in the former case for a GaAs-QD biexciton (XX), and $66\,$ps for an In(Ga)As XX. 

In this work, we report the fabrication of deterministic QD-hCBG with a possible spatial integration precision of  $< 15$\,nm and accuracy of $< 20\,$nm, enabling high Purcell enhancement exceeding 25, which is the highest observed Purcell enhancement in deterministically fabricated hCBG cavities reported so far, resulting in $T_1$ of $<27\,$ps for an InAs QD.

Exploring the impact of the high Purcell-enhancement on the quantum-optical properties of the emitted photons reveals photon-indistinguishabilities of up to 81\% and 96\% under p-shell and s-shell resonant excitation, respectively. Moreover, the strong Purcell enhancement results in improved temperature-robustness in two-photon interference experiments up to 30~K. Finally, we demonstrate fast coherent optical driving of the fabricated devices up to clock-rates of 1.28 GHz with minuscule effects on the multi-photon emission probability and indistinguishability.

\section{Deterministic fabrication}
\label{sec:deterministic_fabrication}

\begin{figure*}[ht]
    \center
	\includegraphics[width=0.95\textwidth]{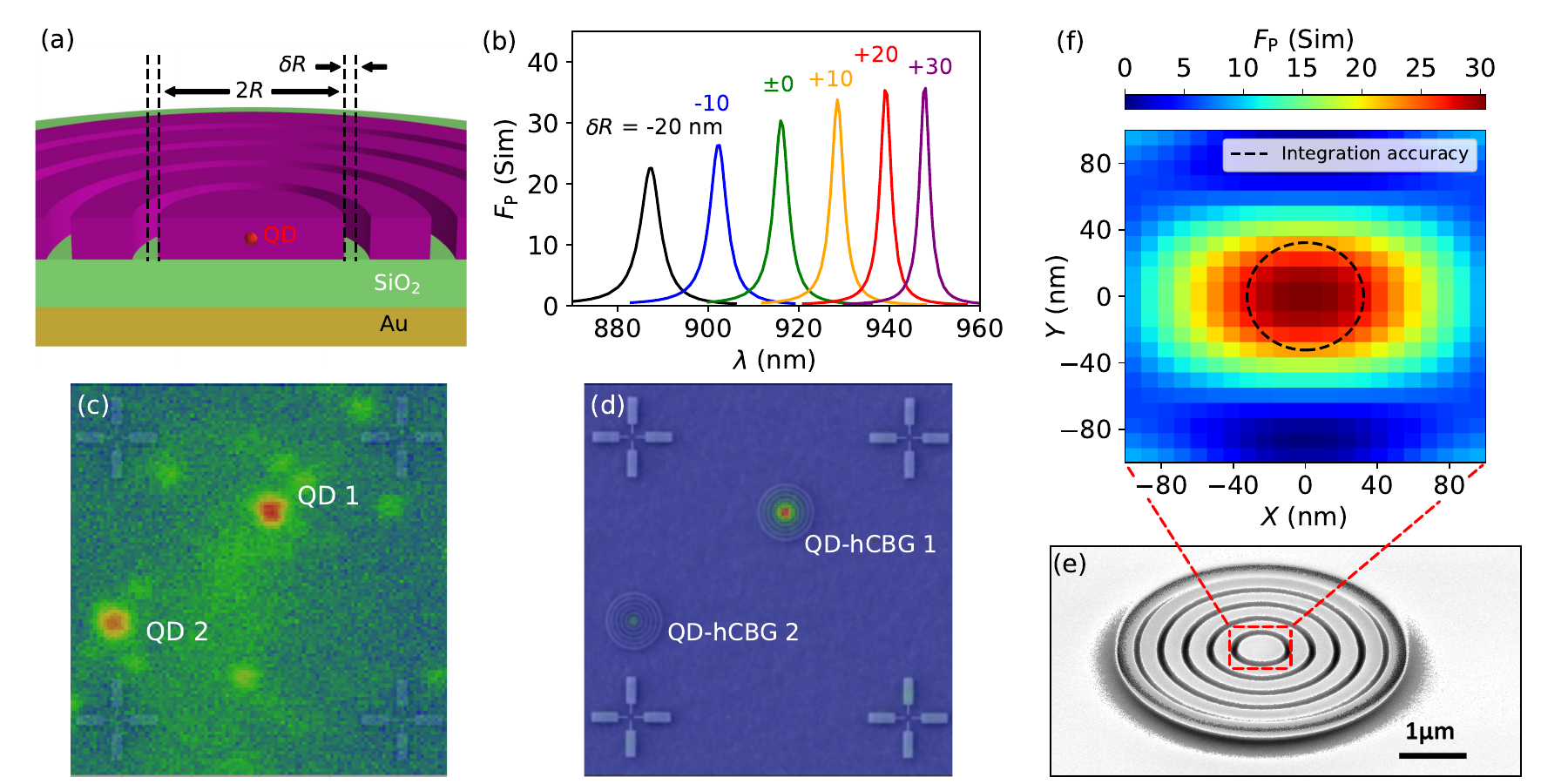}
	\caption{(a) Schematic depiction of a QD-hCBG cavity. $R+\delta{R}$ represent a variation in size of the CBG's central disc. (b) FEM-simulated spectral dependency of the Purcell factor $F_\mathrm{P}$ of an hCBG cavity with $\delta{R}$ variation. (c) Cathodoluminescence (CL-) map of a QD-containing hybrid flip-chip substrate with fabricated markers. Two QD emission patterns are indicated. (d) CL-map in (c) after the marker-aligned fabrication of hCBG cavities at the determined QD-positions. (e) Scanning Electron Microscopy (SEM) image of a fabricated QD-hCBG cavity. (f) Simulated $F_\mathrm{P}$ for one of the two orthogonal polarisations plotted as heatmap for up to $\pm$100~nm spatial mismatch of QD and hCBG cavity. The dashed ellipse corresponds to the estimated spatial integration accuracy $\sigma_\mathrm{spatial}\sim32$\,nm.}
	\label{fig:figure1}
\end{figure*}

A schematic of an hCBG cavity structure with embedded QD and situated on a SiO$_2$ spacer and gold back-reflector is depicted in Figure~\ref{fig:figure1}(a). The radius $R$ of the hCBG's inner disc as well as a controlled variation of this parameter $\delta{R}$ is indicated. Fig.~\ref{fig:figure1}(b) shows the influence of changes in $\delta{R}$ on the simulated Purcell factor $F_\mathrm{P}$ based on Finite Element Method (FEM) simulations, with an expected spectral shift of maximum $F_\mathrm{P}$ to longer (shorter) wavelengths with larger (smaller) $R+\delta{R}$~\cite{yao_design_2018,rickert_optimized_2019}. The target $R$ for the QD-hCBG cavities used in this work is $R=360$~nm, with a variation in $R+\delta{R}$ of -20 to +30 nm shift covering a spectral range of Purcell enhancement from 885-950~nm.  Purcell enhancement $>20$ to $>35$ is possible, while the reduced maximum $F_\mathrm{P}$ as well as the spectrally broader range of Purcell enhancement at smaller $R+\delta{R}$ values stems from a reduced quality factor $Q$ due to weaker mode confinement. Further details on the FEM simulations and hCBG parameters are given in the Supplementary Information (S.I.)~S1.

The hCBG cavities are fabricated based on molecular beam epitaxy (MBE) grown InAs QDs on semi-insulating (100)-GaAs substrates using an In gradient method for low density QDs emitting between 900-940~nm~\cite{shang_proper_2016,chen_telecommunication_2016}. The QDs are embedded in a thin GaAs membrane on two separated etch stop layers and is flip-chip bonded and chemically etched to form hybrid GaAs/SiO$_2$/Au substrates (see S.I. S2). As a next step, marker arrays are fabricated on the substrate. The deterministic QD-hCBG integration on these hybrid substrates is based on a low-temperature, marker-based cathodoluminescence (CL)~\cite{rodt_correlation_2005} mapping process in an electron beam lithography (EBL) system equipped with a cryostat and a CL unit and a following EBL process (acceleration voltage $U_\mathrm{B}$=20~kV) at room temperature aligned to the marker arrays. Details on the deterministic fabrication can be found in S.I. S3.

A CL map with indicated QD signatures can be seen in Fig.~\ref{fig:figure1}(c), superimposed with the simultaneously recorded SEM image with markers before the fabrication of the hCBG cavities. The recorded map size is 29$\times$29~$\mu$m$^2$. The same area with deterministically fabricated cavities is shown in Fig.~\ref{fig:figure1}(d). An SEM image of a fabricated cavity is shown in Fig.~\ref{fig:figure1}(e). 

Before the integration, the QD emission patterns span several $\mu$m in diameter due to the large distance excited carriers can diffuse in the substrate (Fig.~\ref{fig:figure1}(c)). We estimate the error of determining the QD's position $\sigma_\mathrm{QD}\sim27$\,nm from its CL-emission by repeated recording of the same CL-maps, and comparing the extracted positions (see S.I. section S4). After the deterministic integration, the QD's emission is confined to the the central disc of the hCBG cavity caused by the interaction with the mode and the finite interaction volume of the insulated disc on the dielectric spacer (Fig.~\ref{fig:figure1}(d)). We furthermore observe that the central position of the emission in the CL map after the integration only varies $<2$~nm, if the spectral region of the QD or of the wetting layer is selected for integration. Thus, rather than the QD-emission profile, the CL-emission map after the integration yields an accurate measure of the hCBG cavity's center position.
We can thus estimate the positioning accuracy $\sigma_\mathrm{hCBG}$ of the hCBG cavity in the EBL step by comparing the extracted hCBG position to the target value. Two reference processes (see S.I. section S4) yielded an average $\sigma_\mathrm{hCBG}\sim18$\,nm, allowing to estimate the overall integration accuracy as $\sigma_\mathrm{spatial} = \sqrt{\sigma_\mathrm{QD}^2+\sigma_\mathrm{hCBG}^2}\sim 32$\,nm.
% Therefore, comparing the QD's emission center extracted from CL maps before the fabrication with respect to the specific map's marker array to the hCBG's emission center after the fabrication to the same marker array allows for straight forward investigation of the spatial integration precision of the process. The difference between the extracted positions before and after the integration for several hCBG cavities of two exemplary processes $A$ and $B$ are plotted in Fig.~\ref{fig:figure1}(f). Alongside, the spatially-resolved maximum $F_\mathrm{P}$ obtained from FEM simulations for $\delta{R}=0~\mathrm{nm}$ for one of the two linear orthogonal polarisations of the hCBG mode is shown. The mean position deviation in $x-$ and $y-$position for Process A (B) are -0.1~nm/13.8~nm (-6.0~nm/30.4~nm) with standard deviations of 10.7~nm/16.0~nm (9.1~nm/14.6~nm). Based on these mean values and standard deviations we conclude that integration precisions $< $15~nm and accuracies $< $20~nm can be achieved with our process approach, well within the area of $F_\mathrm{P} >20$.

The obtained accuracy-values with the presented marker-based CL technique here are better than the integration performance reported in a recent systematic study that compared marker-based PL- and in-situ CL integration techniques~\cite{madigawa_assessing_2024}. Reasons for the superior performance of the marker-based CL approach compared to the marker-based PL-approach might be a more accurate determination of the marker positions by the electron beam, but could also originate from the marker-aligned EBL process, e.g. by the considerably higher acceleration voltages used in Ref.~\cite{madigawa_assessing_2024}, which might lead to lower marker contrasts during the alignment, resulting in worse alignment precision. The higher accuracy-values compared to the in situ CL integration technique, which is also carried out at cryogenic temperatures, might originate from the fact, that in case of a present temperature-induced drift during the mapping, the recording of the marker positions can compensate for this drift. It is however also important to note that the sample properties like CL emission pattern and brightness of the QDs can affect the accuracy of determining the QD's position from the CL maps, and direct comparison between the in-situ CL technique in Ref.~\cite{madigawa_assessing_2024} and the marker-based CL integration presented here would need to be carried out for the same sample to allow for comparison.

The estimated integration accuracy using the marker-based CL method in this work is plotted in Fig.~\ref{fig:figure1}(f) together with the spatially-resolved maximum $F_\mathrm{P}$ obtained from FEM simulations for $\delta{R}=0~\mathrm{nm}$ for one of the two linear orthogonal polarisations of the hCBG mode.
We point out that the mode orientation in the fabricated structures might be rotated in respect to the one shown in Fig.~\ref{fig:figure1}(f). The mode orientation here can be regarded as a worst-case-scenario, which still predicts that experimental $F_\mathrm{P}>20$ are well within reach for the presented integration approach.

\begin{figure*}[t]
    \center
	\includegraphics[width=0.95\textwidth]{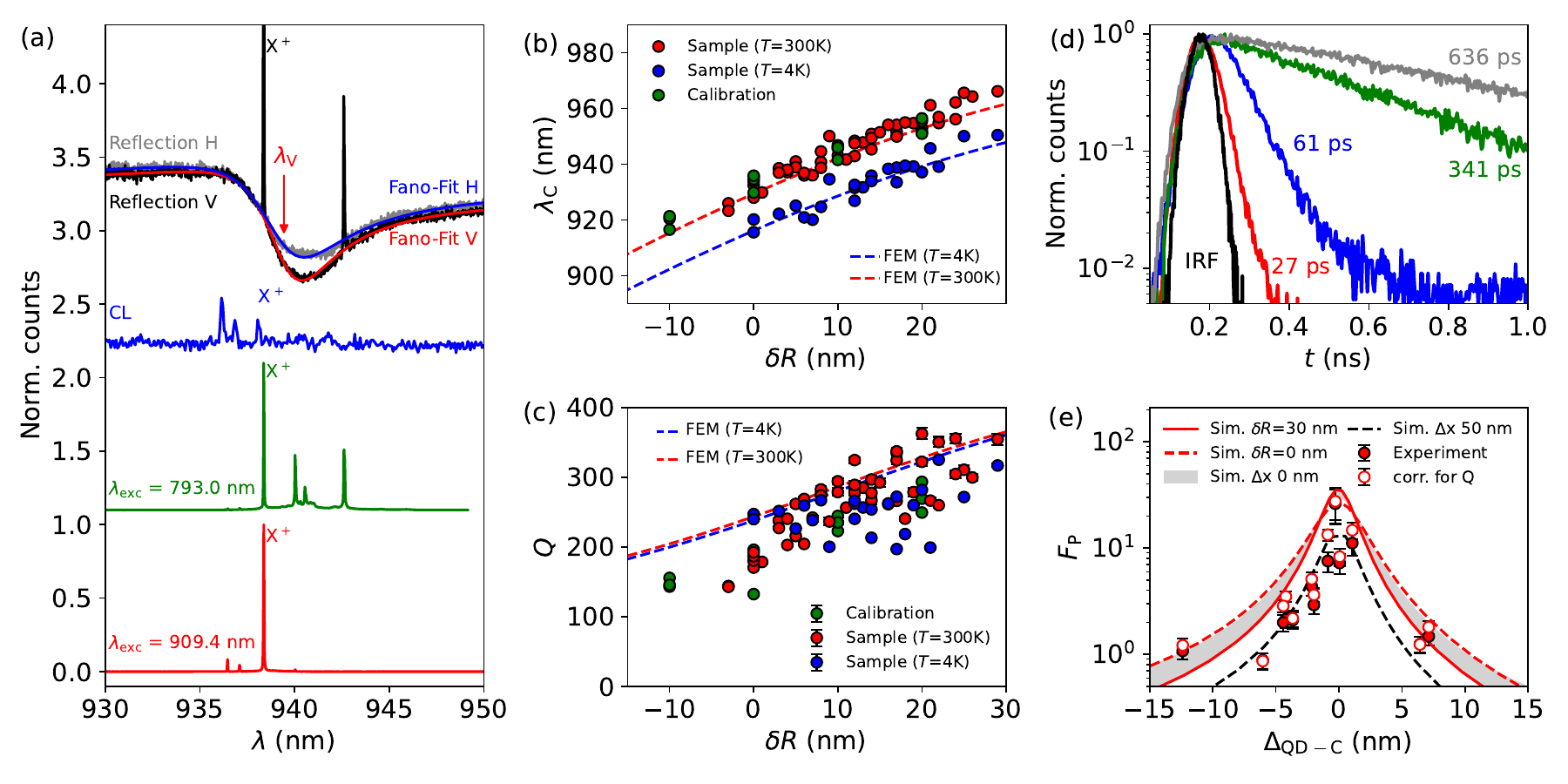}
	\caption{(a) QD-hCBG spectra obtained under white-light excitation, CL spectra before the integration and PL spectra in above-band and quasi-resonant excitation. (b) and (c) Experimental and theoretical cavity mode wavelength $\lambda_\mathrm{C}$ and experimental $Q$-factor at low and room temperature for deterministically integrated hCBGs and prior calibration process for fabricated $\delta{R}$ of the hCBG. (d) Time-resolved measurements in logarithmic scale and extracted $T_1$-times for several X$^+$ transitions in deterministic QD-hCBG cavities with varying spectral detuning, alongside the instrument response function (IRF). The extracted $T_1$-times by exponential fitting are indicated. (e) Experimental $F_\mathrm{P}$ values obtained from $T_1$-times with varying spectral detuning, as well as corrected for reduced Q-factors in the experiment for easier comparison to the simulations. For ideal emitter placement, the red solid line corresponds to the $F_\mathrm{P}$ of a cavity with $\delta{R}=30$\,nm, while the red dashed line shows the $F_\mathrm{P}$ for $\delta{R}=0$\,nm. The gray shaded curve indicates the theoretical $F_\mathrm{P}$ for $\delta{R}$-values in-between these two extrema for the investigated hCBGs. The black dashed line represents Purcell enhancement for an emitter that is spatially misaligned to the hCBG mode by 50~nm.}   
	\label{fig:figure2}
\end{figure*}

With sufficient spatial matching achieved using the marker-based CL-approach described above, we focus now on the spectral overlap between QD and cavity mode. Figure~\ref{fig:figure2}(a) shows spectrally resolved data of an exemplary QD-hCBG cavity for varying excitation methods, including the aforementioned CL-spectrum (blue), micro-photoluminescence ($\mu$PL)-spectra obtained under pulsed laser-excitation with two different excitation energies (green, red), and polarization-resolved reflection spectra (black) obtained by illumination with a broadband white-light-source. The $\mu$PL and reflection spectra were obtained by transferring the sample to a confocal $\mu$PL setup and cooling it to 4~K. Detailed information on the optical setup can be found in S.I.~S5.

The cavity mode's spectral position can be extracted from the white-light reflection measurements by fitting the data with a Fano function~\cite{galli_light_2009}, as indicated in red (blue) for the V- (H-)polarized cavity mode. The extracted V-polarized cavity mode wavelength $\lambda_\mathrm{V}$ from the Fano-fit is indicated. Apart from observing the cavity mode dips in the reflected signal, the white light is already sufficient to excite specific states of the embedded QD. The target QD emission line (the positive trion (X$^+$) transition) is indicated. The second line is assigned to the neutral exciton. The assignment of the QD states is based on excitation power and polarization dependent measurements, as shown in S.I.~S6.

The X$^+$ emission line can also be found in the CL spectrum before the hCBG cavity fabrication (blue line in Fig.~\ref{fig:figure2}(a)), with additional lines blue-shifted from the trion, which we assign to higher charged states like X$^{2+}$~\cite{shang_c2v_2020}. The spectra plotted in green and red show PL-spectra under above-band excitation (excitation wavelength $\lambda_\mathrm{exc}$=793~nm) and quasi-resonant excitation ($\lambda_\mathrm{exc}$=909.4~nm), respectively. We note, that for this specific wafer material, the neutral states require longer integration times and higher excitation power, to see their signatures in the CL signal, while they appear readily in the off-resonant PL-spectrum. We attribute this to the different excitation mechanisms for CL and PL, with the former exciting (positively) charged states more effectively for the present sample, which stems from the specific given doping background. We refer to Ref.~\cite{rodt_correlation_2005}, which also showed present positive trion signatures from CL, but with more prominent neutral states. Another indicator for noticeable positive background doping in the given sample is that photoluminescence excitation scans (see. S.I.~S7) show that the X$^+$-state is most effectively excited under quasi-resonant conditions. 

It is apparent from Fig.~\ref{fig:figure2}(a) that there is a small, but noticeable shift in the X$^+$ wavelength observed from CL and PL spectra of about 1~nm. This originates from observed Stark-tuning due to charge-accumulation in the thin GaAs membrane on insulator during the CL-mapping, and can also be partly caused by the spectrometer alignment of the CL-system compared to the used PL-setup. While the influence of Stark-tuning can in the future be optimized by doped-samples substrates or other means of electrical grounding, we estimate that the determination of the targeted X$^+$ wavelength in this work is limited to $\geq0.5$~nm.

Since the used substrates in this work do not allow for electrical or strain tuning, the spectral QD-mode matching is fixed by the aforementioned target X$^+$ wavelength and the hCBG cavity dimensions. However, as shown in Fig.~\ref{fig:figure1}(b), variations of the inner disc size allow in principle a precise setting of the hCBG's cavity mode wavelength by 1.3~nm per 1~nm $\delta{R}$ variation. Fig.~\ref{fig:figure2}(b) shows mode wavelengths extracted from reflection measurements at room and low temperature for different variations of the central disc size, emphasizing that a 1 nm $\delta{R}$ can be achieved, and is in good agreement with FEM simulations (dashed lines). 

Fig.~\ref{fig:figure2}(c) shows the corresponding cavity $Q$-factor extracted from the width of the mode dip in the reflection measurements shown in (b), further confirming that the fabricated hCBGs reach the theoretically possible $Q$ (and therefore also the expected achievable $F_\mathrm{P}$) as indicated by theory. The data points in green correspond to a fabrication process on the sample prior to the fabrication of the deterministic cavities, to confirm the assumed refractive indices and calibrate the $\delta{R}$-dependency for the sample at hand. This step is necessary, because we found that even for sample pieces from the same wafer but at different wafer-positions, the fluctuations in thickness are enough for a systematic offset of the theoretically expected mode wavelength.

Despite this initial calibration, it is however apparent that fabricated cavities with nominally identical $\delta{R}$-value show wavelength variations of up to 2~nm, and some $Q$ factors deviate more than 20\% from the theoretical $Q$. Furthermore, we observe a fabrication-induced polarisation-dependent mode-splitting between 0 and 3~nm due to ellipticity of the central disc. We therefore estimate the overall spectral matching accuracy to be $\geq$1~nm at best, if we fabricated the theory-predicted $\delta{R}$-attuned cavity on the selected QDs. In order to investigate the effect of different $F_\mathrm{P}$ on the quantum-optical properties, and to increase the probability of ideal spectral matching, we fabricate cavities with varying $\delta{R}$ for groups of QDs with similar X$^+$ wavelengths.

Fig.~\ref{fig:figure2}(d) shows time-resolved measurements of the photon arrival time distribution in semi-logarithmic scaling for spectrally filtered emission from the X$^+$ transitions under pulsed p-shell excitation for QD-hCBGs with varying spectral detuning. The $T_\mathrm{1}$-times are extracted by exponential fits, and are indicated for each curve. The $T_1$ time without a hCBG cavity, i.e. in the surrounding membrane and therefore with a $F_\mathrm{P}=1$ is determined by averaging over several X$^+$ decays measured QDs in the surrounding membrane area under similar excitation conditions to be $680(111)$~ps (see. S.I. S7). The extracted $T_1$-times displayed in Fig.~\ref{fig:figure2} correspond to experimentally measured Purcell enhancements of up to $F_\mathrm{P}>25$ compared to the surrounding membrane in the case of the shortest $T_1$ of $26.9(9)$~ps (red curve), which is still within the resolution of the employed superconducting nanowire detector system, as indicated by the instrument response function (IRF).

The corresponding experimentally determined $F_\mathrm{P}$ extracted for different spectral detunings $\Delta_\mathrm{QD-C}$ between QD transition and cavity mode  of each of the investigated hCBGs is shown in Fig.~\ref{fig:figure2}(e). The error margins of the experimentally determined $F_\mathrm{P}$ originate from the standard deviation of the distribution of observed planar $T_1$ times for reference. The gray shaded area indicates the different simulated $F_\mathrm{P}$ due to the employed varying inner disc size of the hCBG cavity to achieve spectral matching to the QDs, assuming no lateral position deviation. The red solid line corresponds to $\delta{R}=30$\,nm, while the red dashed line is the resulting $F_\mathrm{P}$ for $\delta{R}=0$\,nm. All other $F_\mathrm{P}$-dependencies of the fabricated cavities lie between these two curves in the shaded area. Since some of the cavities exhibited reduced Q-factors compared to the simulations (see Fig.~2(b)), the open circles in Fig.~2(e) show the experimentally measured $F_\mathrm{P}$ corrected for the reduced Q-factors by $F_\mathrm{P}^\mathrm{corr}=F_\mathrm{P}/(Q_\mathrm{exp}/Q_\mathrm{sim})$, resulting in an increase between 1\% and 45\% for better comparison to the simulated Purcell enhancement. With the effects of the reduced experimental $Q$ taken into account, the remaining deviations of the measured Purcell enhancement compared to the simulations are likely to originate from a lateral displacement of the QD from the center of the cavity. The dashed black line represents the theoretical Purcell enhancement for a spatial mismatch of $\Delta{x}=50$~nm between QD and hCBG mode. The fact, that the majority of the experimental data lies above this line indicates the high degree of spatial integration precision with the presented approach.

The achieved $T_1$-time in this work is close to the shortest reported In(Ga)As $T_1$-times in literature in any kind of photonic resonator to-date, which, to the best of our knowledge, are $(22.7\pm0.9)\,$ps, achieved in a photonic crystal cavity with $F_\mathrm{P}$ of $\sim$43~\cite{liu_high_2018}, and $(22.6\pm0.7)$\,ps, recently surpassing this $T_1$-value in a metal-clad nanopillar with $F_\mathrm{P}$ of $\sim$36~\cite{chellu_purcell-enhanced_2024}. We note that the target $F_\mathrm{P}$ of around 30 of the hCBGs presented here could readily increased to 50 and more by modifications of the dielectric spacer and grating parameters~\cite{rickert_optimized_2019}.

\section{(Quasi-)resonant excitation}
\label{sec:T1_and_quantumoptics}

Before turning our focus to the influence of strong Purcell enhancement on the indistinguishability of the emitted photons by the QD-hCBG cavities with varying $T_1$-times, we investigate the influence of the excitation method on the quantum-optical properties. For details regarding the experimental setup to acquire the following data, including equipment and spectral filtering conditions, we refer once again to S.I.~S5.

\begin{figure*}[ht]
    \center
	\includegraphics[width=0.95\textwidth]{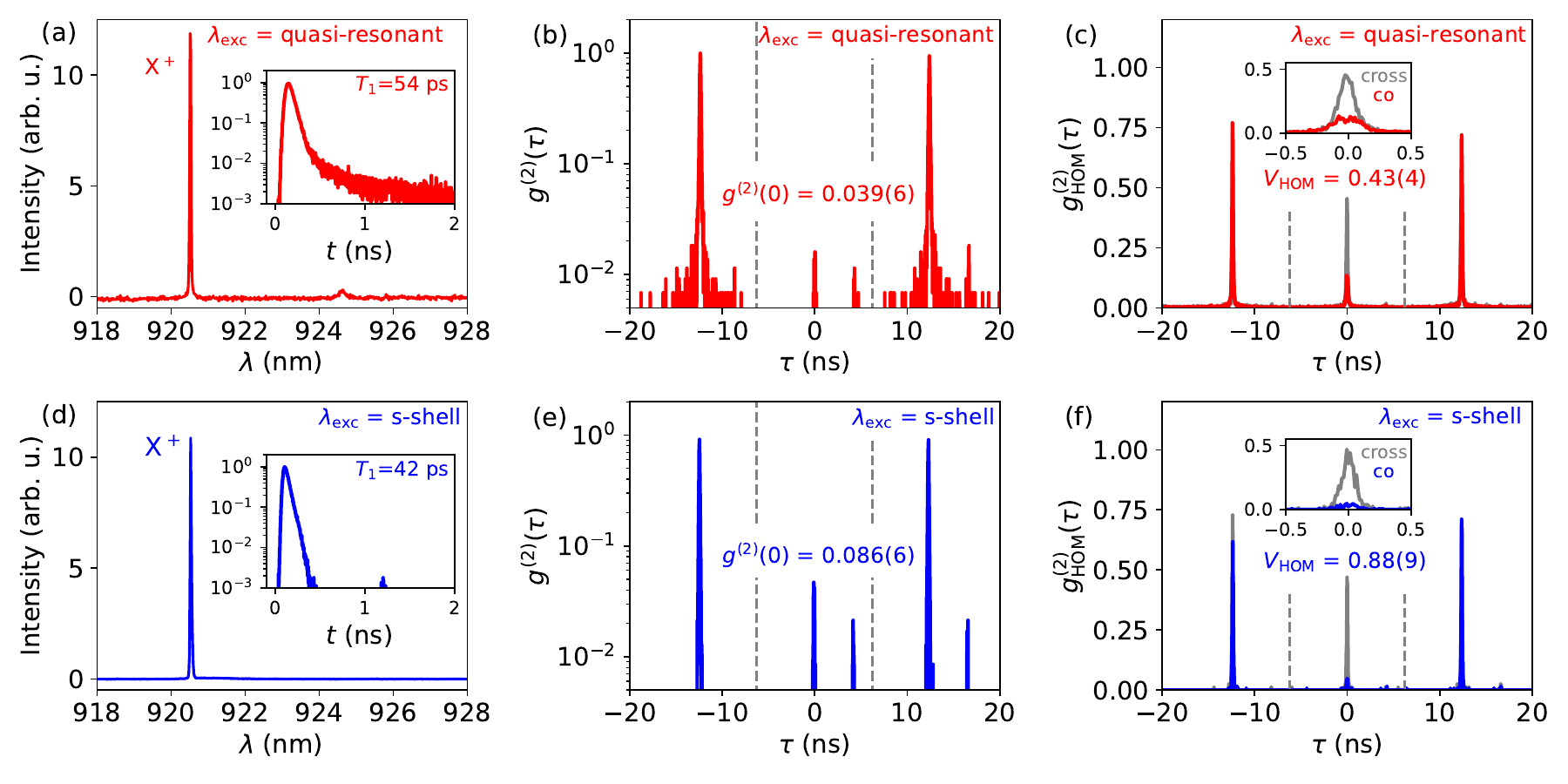}
	\caption{(a) $\mu$PL-spectrum of a QD-hCBG cavity under quasi-resonant excitation and corresponding time-resolved measurement as inset. (b,c) Time-resolved second-order-auto-correlation measurements in a Hanbury-Brown and Twiss (HBT) configuration $\mathrm{g}^{(2)}(\tau)$ and two-photon-interference measurements in Hong-Ou-Mandel (HOM) configuration $g^{(2)}_\mathrm{HOM}(\tau)$ for emitter and conditions as in (a). (d) $\mu$PL-spectrum and time-resolved measurement (inset) for the same QD-hCBG cavity as in (a), but for strictly resonant (s-shell) excitation of the X$^+$ state with a $\pi$-pulse. (e,f)  $\mathrm{g}^{(2)}(\tau)$ and $g^{(2)}_\mathrm{HOM}(\tau)$ for the same conditions as in (d).}
	\label{fig:figure3}
\end{figure*}

Figure~\ref{fig:figure3}(a) shows the emission spectrum of a QD-hCBG cavity with its X$^+$ emission line at $\lambda_\mathrm{X^+}$=920.5~nm under quasi-resonant  excitation at $\lambda_\mathrm{exc}$=891.5~nm at a higher QD-shell (see S.I.~S7.) The corresponding time-trace and extracted $T_1$=54(1)~ps ($F_\mathrm{P} = 12.6(2.3))$ for this case are shown as an inset. Besides the fast $T_1$-decay, the time-trace exhibits a second, slower decay of several hundred ps under these excitation conditions.

The results of time-resolved second-order-auto-correlation measurements in a Hanbury-Brown and Twiss (HBT) configuration $\mathrm{g}^{(2)}(\tau)$ under these excitation conditions can be seen in Fig.~\ref{fig:figure3}(b). A $\mathrm{g}^{(2)}(0)$-value of 0.039(0.6) obtained from comparing the integrated peak at $t=0$, to the average area of neighbouring peaks. The integration window of 12.5~ns is indicated, corresponding to the full laser repetition rate. 

Two-photon-interference measurements (TPI) in Hong-Ou-Mandel (HOM) configuration $g^{(2)}_\mathrm{HOM}(\tau)$ for this QD-hCBG cavity for consecutively emitted photons with a temporal delay of $\delta{t}=12.5$~ns were recorded. We estimate a spectral filtering of about 100~$\mu$eV resulting from the employed spectrometer grating together with the spatial collection of the single-mode fiber incoupling, resulting in no filtering of the X$^+$'s zero phonon line (ZPL), but filtering of its phonon side band to some extent. 

The $g^{(2)}_\mathrm{HOM}(\tau)$-measurement under these conditions for quasi-resonant excitation are shown in Fig.~\ref{fig:figure3}(c). The histograms in red and grey correspond to co-polarized and cross-polarized measurement conditions. From the peak areas around $t$=0~ns of both measurements, a raw, i.e. uncorrected visibility of $V_\mathrm{HOM}=1-A_\mathrm{co}/A_\mathrm{cross}=0.43(4)$ is extracted. This visibility can be corrected for the finite $g^{(2)}(0)$ and HOM-setup limitations~\cite{zhai_quantum_2022,santori_indistinguishable_2002} to $V^{\mathrm{corr}}_\mathrm{HOM}\approx V_\mathrm{HOM} + 2g^{(2)}(0) = $0.51(4). 

The identical QD-hCBG cavity can also be excited using strictly resonant excitation (s-shell) of the X$^+$ state to achieve higher TPI-visibilities due to higher degrees of coherence of consecutively emitted photons. The resulting spectrum for a pulse area of $\pi$ is shown in Fig.~\ref{fig:figure3}(d). To filter the resonant excitation laser light, a cross-polarization configuration~\cite{nick_vamivakas_spin-resolved_2009} is used. In addition, we employ weak above band support using a strongly attenuated continuous-wave laser at 730\,nm, which was necessary to achieve s-shell excitation for several (but not all) QDs on this sample. We attribute this observation to the stabilization of the charge environment in case of specific quantum emitters.

Note that suppressing the reflected excitation laser to resonantly excite QDs in complex structures such as hCBG cavities is not straightforward due to the increased laser scattering by the photonic structure~\cite{von_helversen_triggered_2022}. Previously, the laser suppression was enhanced by fabricating elliptical hCBG structures~\cite{wang_towards_2019}, while the devices fabricated in this work are symmetrical. Here, the necessary laser suppression is achieved by a low required excitation power due to the high hCBG in-coupling efficiency, so that typical $\pi$-powers are of the order of 1~nW. The coherent excitation is confirmed by the observation of Rabi rotations as shown in S.I.~S8.

The resulting time-trace for the X$^+$ is shown in the inset. The $T_1$ is slightly reduced to about 42~ps, and also no additional bi-exponential decay is observed in contrast to the quasi-resonant excitation conditions. This indicates that the bi-exponential decay-behavior stems from a meta-stable decay level which is involved during the relaxation of the charge-carriers into the s-shell before recombination, since it is not observed, if the s-shell is directly populated under strictly resonant excitation.  

In contrast to the aforementioned quasi-resonant excitation, the laser pulse was spectrally filtered using a folded 4f pulse-shaper to improve the spectral matching between laser and X$^+$ emission, resulting in an increase of the laser-pulse width in the time domain from about 2\,ps to 8\,ps. While limiting $g^{(2)}(0)$ to 0.086(6) under RF excitation (cf., Fig.~\ref{fig:figure3}, these excitation conditions were evaluated as a good trade-off between re-excitation of the quantum dot and laser-leakage in the collection path, due to scattered laser-light from the fabricated hCBG cavity, limiting the spatial suppression.

Despite the noticeable multi-photon events, the raw indistinguishability under RF excitation is vastly improved compared to quasi-resonant excitation. The respective $g^{(2)}_\mathrm{HOM}(\tau)$-measurement is shown in Fig.~\ref{fig:figure3}(f). The extracted raw visibility is $V_\mathrm{HOM}=0.88(9)$, obtained by integrating again over the full laser repetition period. We point out, that using $V_\mathrm{HOM}+2g^{(2)}(0)$ as correction would yield an unrealistic corrected visibility greater than unity. Previous work~\cite{ollivier_hong-ou-mandel_2021} showed that different potential causes of elevated $g^{(2)}(0)$-values have to be taken into account for visibility correction - with e.g. the effects of stray laser light being more accurately described by a correction of $V^\mathrm{corr}_\mathrm{HOM} = V_\mathrm{HOM}+B g^{(2)}(0)$ using a factor $1\leq {B} \leq{2}$. A conservative visibility correction using $B=1$ yields a corrected HOM visibility under RF excitation of $V^{\mathrm{corr}}_\mathrm{HOM} \geq 0.96$ for the measurement in Fig.~\ref{fig:figure3}. This value is among the highest reported indistinguishabilities at this time delay and RF excitation for In(Ga)As QDs at these wavelengths in literature~\cite{ding_high-efficiency_2023,tomm_bright_2021,somaschi_near-optimal_2016,liu_high_2018}. This is even more noteworthy, since the employed QD-hCBG devices in this present work have no active control of the charge environment, and the observed indistinguishability is therefore expected to be still decreased due to charge fluctuation in the QD's vicinity.

We further point out that both the $g^{(2)}(\tau)$ and $g^{(2)}_\mathrm{HOM}(\tau)$ measurements shown in Fig.~\ref{fig:figure3} exhibit small amounts of signal from a back-reflection towards the detection system most likely caused at a fiber-to-fiber mating. This is especially apparent in Fig.~\ref{fig:figure3}(e) but also visible in the inset of Fig.~\ref{fig:figure3}(d). All obtained $g^{(2)}(0)$ and $V_\mathrm{HOM}$ are not corrected for that, i.e. no temporal filtering has been applied. 

Besides the very high growth-quality of the employed substrates, we attribute the observed high indistinguishability to the strong Purcell enhancement, and short $T_1$ time. We will investigate the influence of varying $T_1$-times on the indistinguishability in the following.

\subsection{\texorpdfstring{$T_1$}{T1} influence on \texorpdfstring{$V_\mathrm{HOM}$}{VHOM}}

\begin{figure*}[ht]
    \center
	\includegraphics[width=0.95\textwidth]{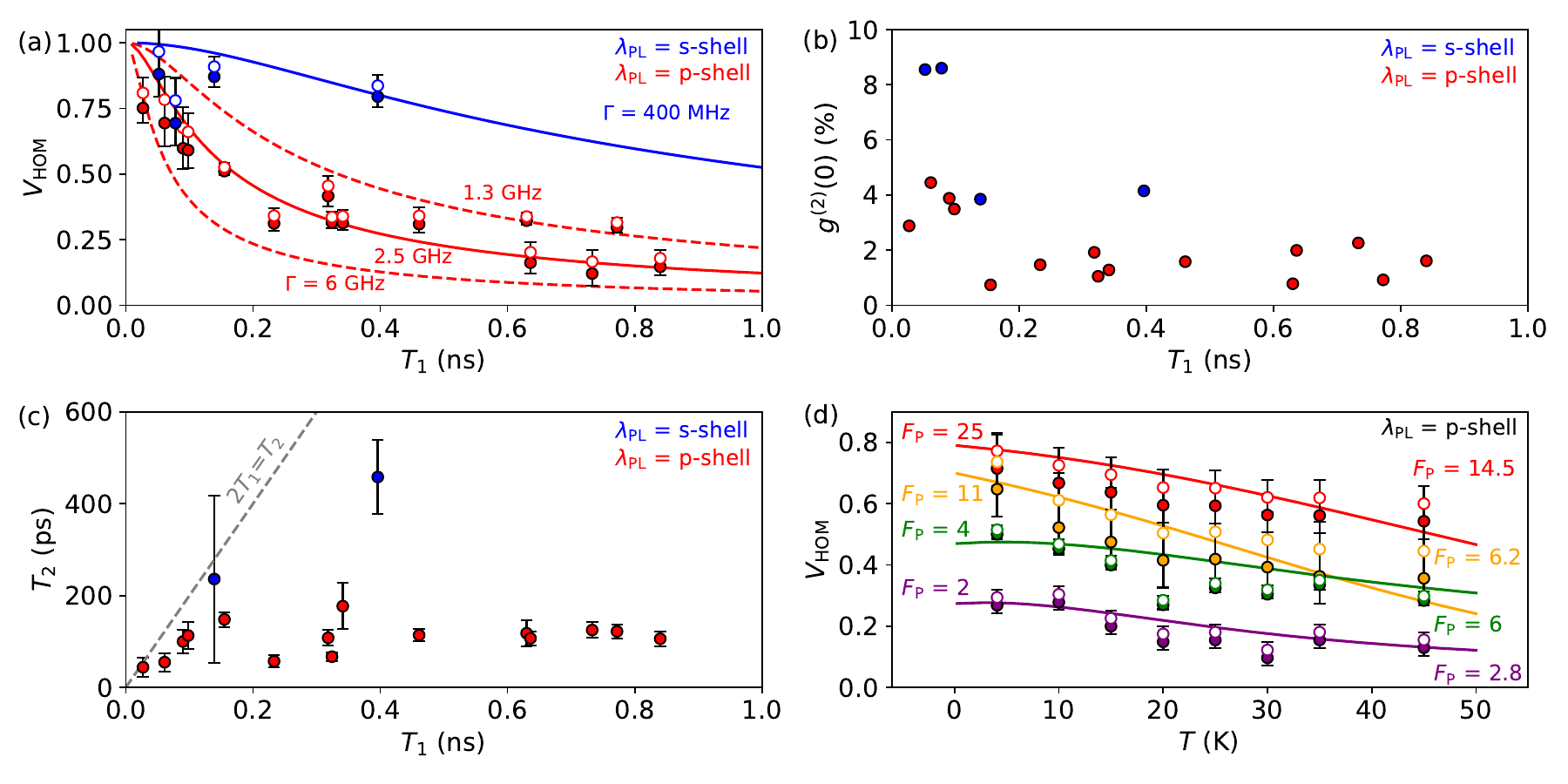}
	\caption{(a) Measured raw $V_\mathrm{HOM}$ (filled circles) and corrected $V^\mathrm{corr}_\mathrm{HOM}$ (open circles) for varying X$^+$ $T_1$ times under p-shell and s-shell excitation. The plotted theory curves are according to equation~(\ref{eq:VHOM_T1}) with respective $\Gamma$-values. (b) Corresponding $g^{(2)}(0)$-values for the data in (a). (c) Corresponding $T_2$-time extracted from fits of the central HOM-dip for the data in (a). (d) Measured raw $V_\mathrm{HOM}$ (filled circles) and corrected $V^\mathrm{corr}_\mathrm{HOM}$ (open circles) for varying X$^+$ transitions with varying $F_\mathrm{P}$ under p-shell excitation. The plotted theory curves are according to equation~(\ref{eq:VHOM_Temp}), taking the $T$-dependent $T_1$-time due to temperature tuning of the embedded QDs into account. The indicated $F_\mathrm{P}$-values are measured at $T=4$\,K and $T=45$\,K for the respective cavities.}
	\label{fig:figure4}
\end{figure*}

Extracted TPI visibilities for varying $T_1$ times are displayed in Figure~\ref{fig:figure4}(a), for p-shell and s-shell excitation of the specific QD-hCBG system. Note that all the measurements were taken with the identical degree of spectral filtering via a monochromator and following single mode fiber in-coupling, as described above. Both the raw and corrected extracted visibilities (latter with $B=2$ for p-shell and $B=1$ for s-shell, see above) are plotted. The error bars correspond to the measured $B\cdot g^{(2)}(0)$-values for the specific QD-hCBGs. The dashed lines are $T_1$-dependent theoretical visibilities based on equation~(\ref{eq:VHOM_T1})~[\cite{gold_two-photon_2014,nawrath_resonance_2021}]:

\begin{equation}
\label{eq:VHOM_T1}
	V_\mathrm{HOM}(T_1,\Gamma) = \frac{A(\Gamma)}{T_1}\exp{\left(\frac{A(\Gamma)^2}{\pi{T_1^2}}\right)}\cdot\mathrm{erfc}\left(\frac{A(\Gamma)}{T_1\sqrt{\pi}}\right)
\end{equation}
with the complementary error-function $\mathrm{erfc()}$ and the factor $A = \sqrt{\ln{2}}/(\sqrt{2\pi }\Gamma)$, including the inhomogeneously broadened QD emission line following a Gaussian distribution with FWHM-linewidth $\Gamma$.

The beneficial effect of strong Purcell enhancement on the HOM visibility is apparent for both kinds of excitation mechanisms, with highest visibility values of $V_\mathrm{HOM}=0.75(6)$ ($V^\mathrm{corr}_\mathrm{HOM}=0.81(6)$) for a QD-hCBG with $T_1=26.9(9)$~ps under p-shell excitation and the aforementioned $V_\mathrm{HOM}=0.88(9)$ ($V^{\mathrm{corr}}_\mathrm{HOM}=0.96(9)$) for a QD-hCBG with $T_1=41.7(2)$~ps under resonant excitation. 
Additionally, the strongly reduced inhomogeneous broadening under RF excitation compared to quasi-resonant excitation is apparent from comparison to theory, and visibilities of $>80\%$ are still observed even for comparably long $T_1$-times of 400~ps ($F_\mathrm{P}\approx 1.7$) under resonant excitation.

To get additional information on the coherence properties of the interfered photons, we extracted $T_2$-coherence times from the central dip in the co-polarized HOM-measurement via an exponential fit, since the width of the dip is related to the coherence time~\cite{legero_time-resolved_2003}. Fig.~\ref{fig:figure4}(c) shows the $T_1$-resolved $T_2$-times for the investigated QD-hCBG cavities in (a). The dashed grey line corresponds to a Fourier-limited $2T_1 = T_2$ relation. 

It is apparent from the data in Fig.~\ref{fig:figure4}(c) that for the cases of significant Purcell enhancement, the observed $T_1$ and $T_2$ times are getting close to this relation even for quasi-resonant excitation, which in turn results in the observed high TPI visibilities. The $T_2$ times extracted for RF excitation are significantly higher, so that the reduction in $T_1$ is already sufficient for close to Fourier-limited photons at moderate $F_\mathrm{P}$. We point out here that for the data points in Fig.~\ref{fig:figure4} with very high visibility, it was not possible to extract $T_2$ from the co-polarized measurement, because the present multi-photon events mask the dip. However, the comparison clearly indicates that the investigated indistinguishability improvement can be related to shortened $T_1$-times, which renders the emitted photons more resilient to present decoherence during the charge-carrier recombination after excitation.

This increased indistinguishability in the presence of decoherence for shortened $T_1$ times is also beneficial, if this decoherence is caused by further inhomogeneous broadening due to phonon interaction~\cite{grange_cavity-funneled_2015,grange_reducing_2017,brash_nanocavity_2023}. Fig.~\ref{fig:figure4}(d) shows extracted TPI visibilities under p-shell excitation for varying degrees of Purcell-enhancement in the temperature range of $\sim 5-45$~K. Fig.~\ref{fig:figure4}(d) shows clearly the beneficial effect of strong Purcell enhancement on the photon indistinguishability for elevated temperatures, with still $V^\mathrm{corr}_\mathrm{HOM} >60\%$ at $T=30-40$~K for $F_\mathrm{P}=25$. This temperature is interesting from a technological point-of-view, since it is reachable with compact Stirling cryocoolers, operable without the need of liquid helium cooling~\cite{schlehahn_stand-alone_2018,musial_plugplay_2020,gao_quantum_2022}.

The solid lines are theoretical TPI visibilities according to
\begin{equation}
\label{eq:VHOM_Temp}
V_\mathrm{HOM}(T, \Gamma) = V_\mathrm{HOM}(T_1,\Gamma)\dfrac{\gamma}{\gamma + \gamma^\ast(T)}
\end{equation}
where $V_\mathrm{HOM}(\Gamma)$ corresponds to equation~(\ref{eq:VHOM_T1}) and accounts for given dephasing due to the quasi-resonant excitation at given $T_1$ and $\Gamma$, and the second factor assumes Markovian-dynamics for the indistinguishability~\cite{bylander_interference_2003,grange_cavity-funneled_2015,grange_reducing_2017}. Here, $\gamma$ represents the Fourier-limited linewidth $\gamma = \hbar/T_1$ and $\gamma^\ast(T)$ being the broadening of the ZPL due to dephasing by scattering from thermal acoustic photons~\cite{grange_reducing_2017}. According to Ref.~\cite{grange_reducing_2017} we approximate $\gamma^\ast(T) = \alpha n(E)(n(E)+1)$ with $n(E)$ being the Bose-Einstein function. For the curves in Fig.~\ref{fig:figure4}, $\alpha$ is set to 3~$\mu$eV and $E$ to 1~meV, similar to previous works~\cite{grange_reducing_2017}. The corresponding $\Gamma$-values for each curve at given $F_\mathrm{P}$ are between 2.5 and 6~GHz, as obtained by comparison of experimental data and theory in Fig.~\ref{fig:figure4}(a).

Note that the increase in temperature causes an energy shift of X$^+$ and a corresponding change in spectral overlap with the cavity mode. The $T_1$-time for all four cavities therefore varies for the different temperatures, resulting in a $T$-dependent $T_1(T)$ and thus also $\gamma$ in eq.~(\ref{eq:VHOM_Temp}). For example, the QD with $F_\mathrm{P} = 25$ showed $T_1<30$~ps at $T=4$~K, and $T_1=45$~ps ($F_\mathrm{P}=14.5$) at $T=45$~K, reducing $V_\mathrm{HOM}(T_1,\Gamma)$ from 0.76 to 0.64. The Purcell enhancement of the two lower $F_\mathrm{P}$ cavities was on the other hand increased at $T=45$\,K, due to tuning the X$^+$ more into resonance with the respective mode. As shown in S.I. section S9, $T_1(T)$ can be approximated with a linear fit for the considered finite temperature change. The steeper drop of $V_\mathrm{HOM}(T)$ for the two high $F_\mathrm{P}$-cavities is primarily caused by the $T$-dependent $T_1$. The reduction of temperature influence on the indistinguishability at these high $F_\mathrm{P}$ is therefore expected to be even higher than what can be seen in Fig.~\ref{fig:figure4}(d), and could be assessed if the QD-hCBG-mode matching could be restored independently from temperature, as was done in previous works e.g. via Stark-tuning~\cite{grange_reducing_2017}. In S.I. section S9, additional calculations are shown if $T_1$ could be kept constant, predicting $V_\mathrm{HOM}>0.7$ for the high $F_\mathrm{P}$-device in p-shell excitation. On the other hand, it is worth noting that the spectrally broad Purcell enhancement of hCBG cavities can act as a stabilizing element for temperature shifts, as indicated here by maintaining comparably high indistinguishability values due to low changes in $T_1$ also in the absence of independent tuning possibilities. Last but not least, S.I. section S9 also shows theoretical predictions of the indistinguishability for higher temperature under coherent s-shell excitation. Assuming inhomogenous broadening of $\Gamma=400$\,MHz as shown in Fig.~\ref{fig:figure4}(a), $V_\mathrm{HOM}>0.9$ are expected for $T=30$\,K.

The observed indistinguishabilities in accordance with theory show the clear benefit of high Purcell enhancement for TPI at elevated temperatures. One should note, that compared to high-Q cavities, e.g., micropillars, the spectrally broad hCBG cavity mode cannot provide a selective enhancement of the ZPL relative to the PSB. This limits the indistinguishability that can be achieved at elevated temperatures without spectral filtering, and the overall efficiency if spectral filtering is applied~\cite{grange_cavity-funneled_2015,iles-smith_phonon_2017}.

\subsection{GHz operation}
\label{sec:GHz}

\begin{figure*}[ht]
    \center
	\includegraphics[width=0.95\textwidth]{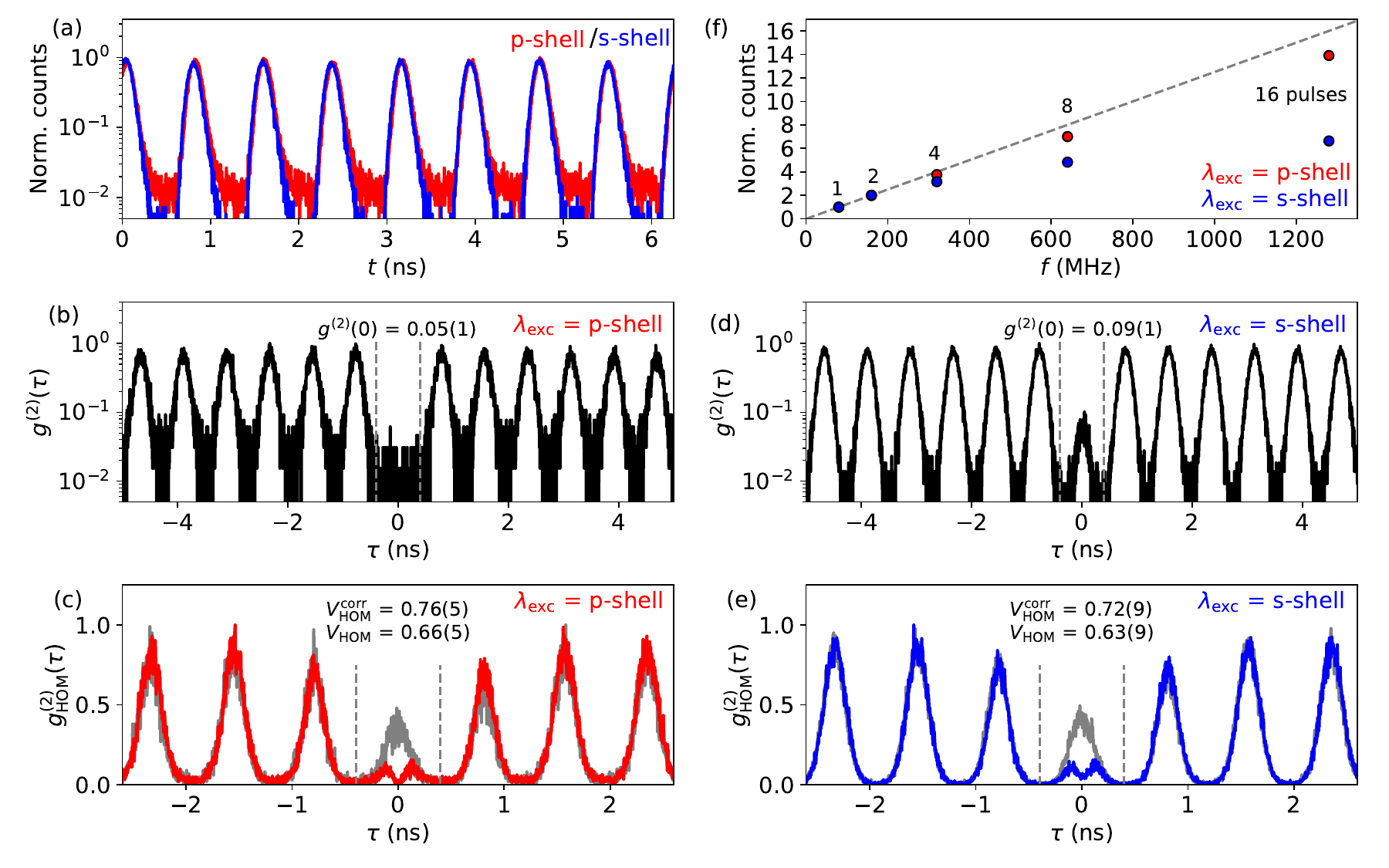}
	\caption{Time-resolved measurements in (quasi-)resonant excitation at 1.28~GHz repetition rate. (a) Lifetime measurements of a QD-hCBGs cavity with $T_1<50$~ps under p-shell and s-shell excitation . (b) Second-order auto-correlation $g^{(2)}(\tau)$-measurement and (c) Two-photon interference $g^{(2)}_\mathrm{HOM}(\tau)$ under p-shell excitation for the QD-hCBG cavity in (a). (d) Second-order auto-correlation $g^{(2)}(\tau)$-measurement and (e) Two-photon interference $g^{(2)}_\mathrm{HOM}(\tau)$ under $\pi$-pulse s-shell excitation for the QD-hCBG cavity in (a). (f) Measured count-rates for an increasing number of excitation pulses (and correspondingly higher excitation rate $f$), normalized by the 1-pulse result for p-shell (red) and s-shell (blue) excitation.}
	\label{fig:figure5}
\end{figure*}

Finally, we demonstrate that the high Purcell enhancement achieved in this work allows for increasing the excitation laser repetition rate into the GHz range. Since we do not have a pulsed excitation laser with repetition rates $>80$~MHz available, we multiply our 80~MHz pulsed laser signal in a self-built sequential rate-doubling setup up to 16-fold, resulting in an excitation pulse separation of $\sim 781$~ps or 1.28~GHz excitation rate. Details on the rate multiplication setup can be found in S.I. S10. Figure~\ref{fig:figure5} shows quantum-optical performances with this GHz rate excitation, for the same QD-hCBG cavity with $T_1$ of 40-50~ps used for the measurements in Fig.~\ref{fig:figure3} at 80~MHz. The $T_1$ time of this QD-hCBG cavity is therefore $>15\cdot{1/f}$.

Fig.\ref{fig:figure5}(a) shows time traces under p-shell and s-shell excitation conditions for this GHz rate. As can be seen, due to the strongly reduced $T_1$-time, the successive peaks are well separated, despite the slower bi-exponential decay component present under quasi-resonant excitation, as discussed previously.

Fig.~\ref{fig:figure5}(b) and (c) show a corresponding $g^{(2)}(\tau)$ and $g^{(2)}_\mathrm{HOM}(\tau)$ measurement under p-shell excitation, yielding $g^{(2)}(0)$= 0.05(1) and $V_\mathrm{HOM}=0.66(5)$ ($V^\mathrm{corr}_\mathrm{HOM}=0.76(5)$) from integrating co-and-cross polarized date over time-windows of 0.781~ns equal to the repetition rate. Comparison to Fig.~\ref{fig:figure3}(a) and (b) shows, that these values at 1.28~GHz excitation rate are nearly identical to driving the QD-hCBG cavity system under 80~MHz excitation rates, therefore confirming that the strong Purcell enhancement enables GHz rate operation without any detrimental effect on the quantum-optical properties for quasi-resonant excitation.

Fig.~\ref{fig:figure5}(d) and (e) show $g^{(2)}(\tau)$ and $g^{(2)}_\mathrm{HOM}(\tau)$ measurements under s-shell 1.28~GHz excitation. While the $g^{(2)}(0)$-value of $0.09(1)$ is in accordance with the observed 80~MHz value, taking re-excitation and imperfect laser-suppression into account, the observed HOM indistinguishability was surprisingly limited: A raw indistinguishability of $V_\mathrm{HOM}=0.63(5)$ ($V^\mathrm{corr}_\mathrm{HOM}=0.72(5)$) was observed, even lower than for the p-shell quasi-resonant excitation. To confirm that the reason is indeed the GHz excitation, we blocked the respective additional excitation pulses while keeping all other conditions (alignment, excitation wavelength, power, filtering) unchanged, and measured the indistinguishability again under 80~MHz s-shell excitation, recovering the $>0.95$-values observed before and described above. We therefore suspect that the increased excitation rate is the cause for the limited $V_\mathrm{HOM}$ in strictly-resonant excitation. 

A further indicator for this is the observed count-rate for different excitation rates, which is shown in Fig.~\ref{fig:figure5}(f). The grey dashed line indicates the expected linear increase in count-rate with pulse number multiplication from 80~MHz (1 pulse) to 1.28~GHz (16 pulses). 
As can be seen, under s-shell excitation (blue dots), the observed count-rate starts to deviate from the expected linear behaviour for $f>$300~MHz (4 pulses, 3.125~ns time window), and reaches only about 25~\% of the expected count-rate at 1.28~GHz or 16 pulses in the 12.5~ns time window. The p-shell GHz excitation (red dots) also shows slightly lower count-rates than the 16-times multiplication, but is much closer to the expected values. 

We suspect this saturation behaviour and the reduced indistinguishability to be linked to the availability of holes to form the X$^+$ ground state. Since the observed time-resolved emission properties, i.e. $T_1$-times do not appear to be influenced by the excitation rate, we suspect that the charge carrier creation prior to the emission is the limiting factor. More concretely, that at excitation rates above 300~MHz, the available time to make additional holes available to keep forming X$^+$-states by optical excitation is too short, limiting the achievable count-rate. 

Likewise, the detrimental effects of high excitation rates on the indistinguishability in s-shell excitation could be explained by the changing charge-environment due to the non-equilibrium conditions for the required excess holes at high excitation rates. Quasi-resonant p-shell excitation on the other hand is energetically above possible hole-states, and might still provide enough additional holes per pulse to limit the effects on the charge-environment at high rates, without influencing the charge environment substantially.  

\section{Conclusions}

In summary, we presented deterministically fabricated QD-hCBG devices with sufficient spatial and spectral matching to achieve Purcell enhancements of up to $F_\mathrm{P}>25$ for a X$^+$ state tuned to the cavity mode, which is the closest value to the expected theoretical performance of these cavities reported so far. We demonstrated the beneficial effects of these strongly reduced $T_1$ times for the generation of indistinguishable photons both for strictly resonant as well as quasi-resonant excitation conditions, and also for application-relevant temperature ranges of $\geq 30$~K. Moreover, the achieved $T_1$-times $<50$~ps enabled excitation rates well beyond the usual 80~MHz driving, with up to 1.28~GHz used in this work, currently just limited by the experimentally achievable excitation rates. While quasi-resonant excitation at these high rates did show no deterioration in the quantum-optical properties, higher excitation rates did show a slight detrimental effect on the two photon indistinguishability for strictly resonant excitation wavelengths. Note however, that this behavior is presumably linked to the type of charged state investigated here and its specific environment and intrinsic doping. Generally speaking, RF should be best suited for GHz excitation rates, since if offers the shortest achievable $T_1$-time.

In order to investigate the origins for this behaviour, further work is needed, considering that very few works have achieved device performances that allow for resonant (optical) GHz driving rates and taking a closer look at the associated emerging physics. 
Since we suspect charge-carrier mechanics to be limiting the current GHz indistinguishability, means of charge control in the form of gated~\cite{buchinger_optical_2023,ma_circular_2024,rickert_high-performance_2023} hCBG cavities appear beneficial to understand and overcome this limitation.  
Our fabrication method is straight-forwardly compatible for such substrates, which would also allow for means of QD-cavity mode matching at elevated temperatures, by tuning the QD emission line back into resonance. First results on photonic ring structures with gated contacts have been recently reported~\cite{wijitpatima_bright_2024}, although without Purcell enhancement so far.  

Since indistinguishabilities close to 0.8 at quasi-resonant wavelengths were already reached here with the Purcell factors achieved in this work, adapted designs could push these values even further, since optimization of the layer structure and CBG dimensions can readily provide Purcell enhancements which are a factor of 2 or 3 higher~\cite{rickert_optimized_2019}. While the associated increased Q-factors would tighten the requirements for spectral QD-mode matching, they would also be beneficial for increasing the indistinguishability and efficiency of the cavities at elevated temperatures~\cite{grange_cavity-funneled_2015} by reducing the amount of photons not emitted into the ZPL.

Lastly, the indistinguishability of currently extensively investigated telecom-QDs~\cite{nawrath_bright_2023,holewa_high-throughput_2024} with so far limited integration accuracy and therefore limited Purcell enhancement would benefit greatly from the high precision integration techniques for spatial and spectral mode matching as shown here. 

\section{Data and Code Availability}
The Data obtained in this work can be provided by the authors upon reasonable request.

\section{Funding}
The authors acknowledge financial support by the German Federal Ministry of Education and Research (BMBF) via the project “QuSecure” (Grant No. 13N14876) within the funding program Photonic Research Germany, the BMBF joint project “tubLAN Q.0” (Grant No. 16KISQ087K), and by the Einstein Foundation Berlin via the Einstein Research Unit “Quantum Devices”. H.L., S.L., H.N., and Z.N. further acknowledge funding by the National Key Technology R\&D program of China (Grant No. 2018YFA0306101) and J.H., S.Rodt, and S.Reitzenstein by the German Federal Ministry of Education and Research (BMBF) through the project QR.X Quantenrepeater.Link (Funding ID 16KISQ014) and German Research Foundation  via project INST 131/795-1 FUGG.

\section{Disclosure}
The authors declare no conflict of interest.

\bibliographystyle{ieeetr}
\bibliography{BibTex_v6}

\setcounter{figure}{0}
\renewcommand{\figurename}{Fig.}
\renewcommand{\thefigure}{S\arabic{figure}}

\onecolumn
\section*{Supplementary Information}
\appendix

\section*{S1: FEM Simulations of hCBGs}
\label{sec:si_FEM}

\begin{figure*}[ht]
    \center
	\includegraphics[width=0.95\textwidth]{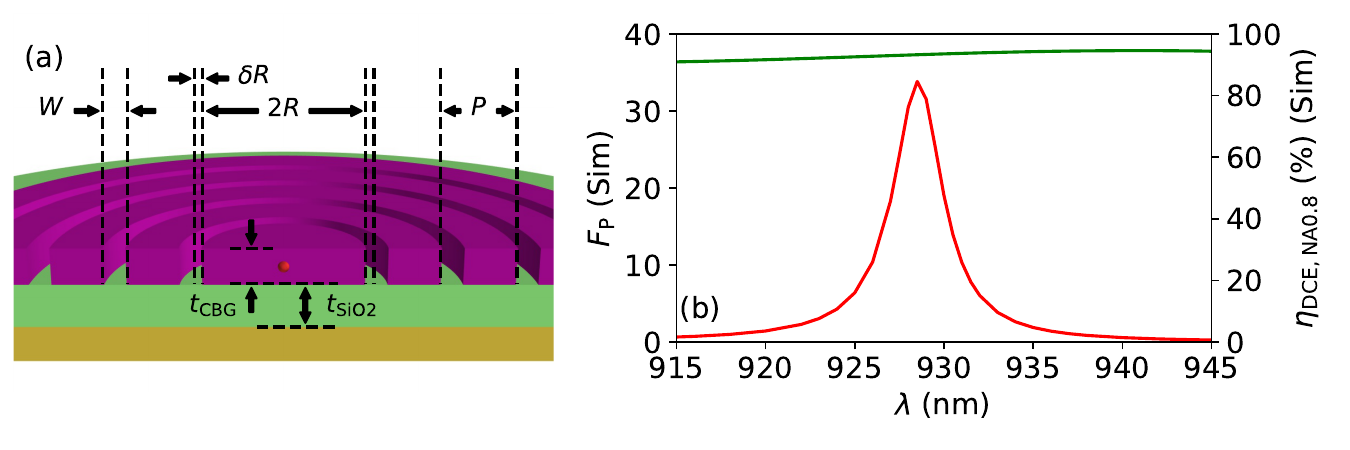}
	\caption{(a) QD-hCBG cavity schematic with indicated design parameters: central disc radius $R$, radius variation $\delta{R}$, grating period $P$, grating gap width $W$, membrane thickness $t_\mathrm{CBG}$ and SiO$\mathrm{2}$ thickness $t_{\mathrm{SiO}_2}$. (b) Simulated Purcell factor $F_\mathrm{P}$ and dipole collection efficiency into NA=0.8 $\eta_\mathrm{DCE,NA 0.8}$ for a hCBG cavity with parameters from Table~\ref{tab:hCBG_parameters} and \ref{tab:indices} and $\delta{R}=10$~nm.}
\label{fig:figureS1}
\end{figure*}

\begin{table}[ht]
\begin{center}
\begin{tabular}{||c c c c c c||}
\hline
$R$ [nm] & $P$ [nm] & $W$ [nm] & $t_\mathrm{CBG}$ [nm] & $t_{\mathrm{SiO}_2}$ [nm] & m$_\mathrm{Rings}$ \\
\hline \hline
360 & 360 & 100 & 170 & 200 & 5 \\
\hline
\end{tabular}
\end{center}
\caption{hCBG cavity design parameters according to Fig.~\ref{fig:figureS1}. The parameter $m_\mathrm{Rings}$ represents the hCBG's number of rings.}
\label{tab:hCBG_parameters}
\end{table}

\begin{table}[ht]
\begin{center}
\begin{tabular}{||c c c c||}
\hline
$n_\mathrm{GaAs, 300K}$ & $n_\mathrm{GaAs, 4K}$ & $n_{\mathrm{SiO}_2\mathrm{,300K}}$/$n_{\mathrm{SiO}_2\mathrm{,4K}}$ & $n_\mathrm{Au}$\\
\hline \hline
3.152 & 3.460 & 1.450 & 0.12 + 6.33$i$\\
\hline
\end{tabular}
\end{center}
\caption{Refractive index parameters used in the FEM simulations.}
\label{tab:indices}
\end{table}

\noindent The FEM simulations presented in this work were performed using the commercial FEM software JCMSuite (JCMwave GmbH, 2024). The QD emitter is simulated by a TE dipole source situated in the centre of the central hCBG disc in both vertical and lateral direction. Further details on the simulation setup can be found in~\cite{rickert_optimized_2019} of the main text. Figure~\ref{fig:figureS1} shows exemplary data from scattering simulations for a typical hCBG cavity design in terms of simulated Purcell factor and fraction of emitted dipole power into Numerical Aperture (NA) of 0.8.  The used parameters for the hCBG cavities are listed in Tables~\ref{tab:hCBG_parameters} and \ref{tab:indices}. 

Fig.~\ref{fig:figure1}(b) displayed in the main text used similar scattering simulations. Fig.~\ref{fig:figure2}(b)(c) used eigenvalue simulations for the same simulation layout. The central mode wavelength was calculated from the real-part of the obtained eigenvalues $\omega$ as $\lambda=(2\pi{c})/\mathrm{Re}(\omega)$, while the Q-factor was obtained from real- and imaginary parts as $Q=\mathrm{Re(\omega)}/(2\mathrm{Im}(\omega))$.

\clearpage
\section*{S2: Substrate fabrication via flip-chip process}
\label{sec:si_flipchip}

\begin{figure*}[ht]
    \center
	\includegraphics[width=0.95\textwidth]{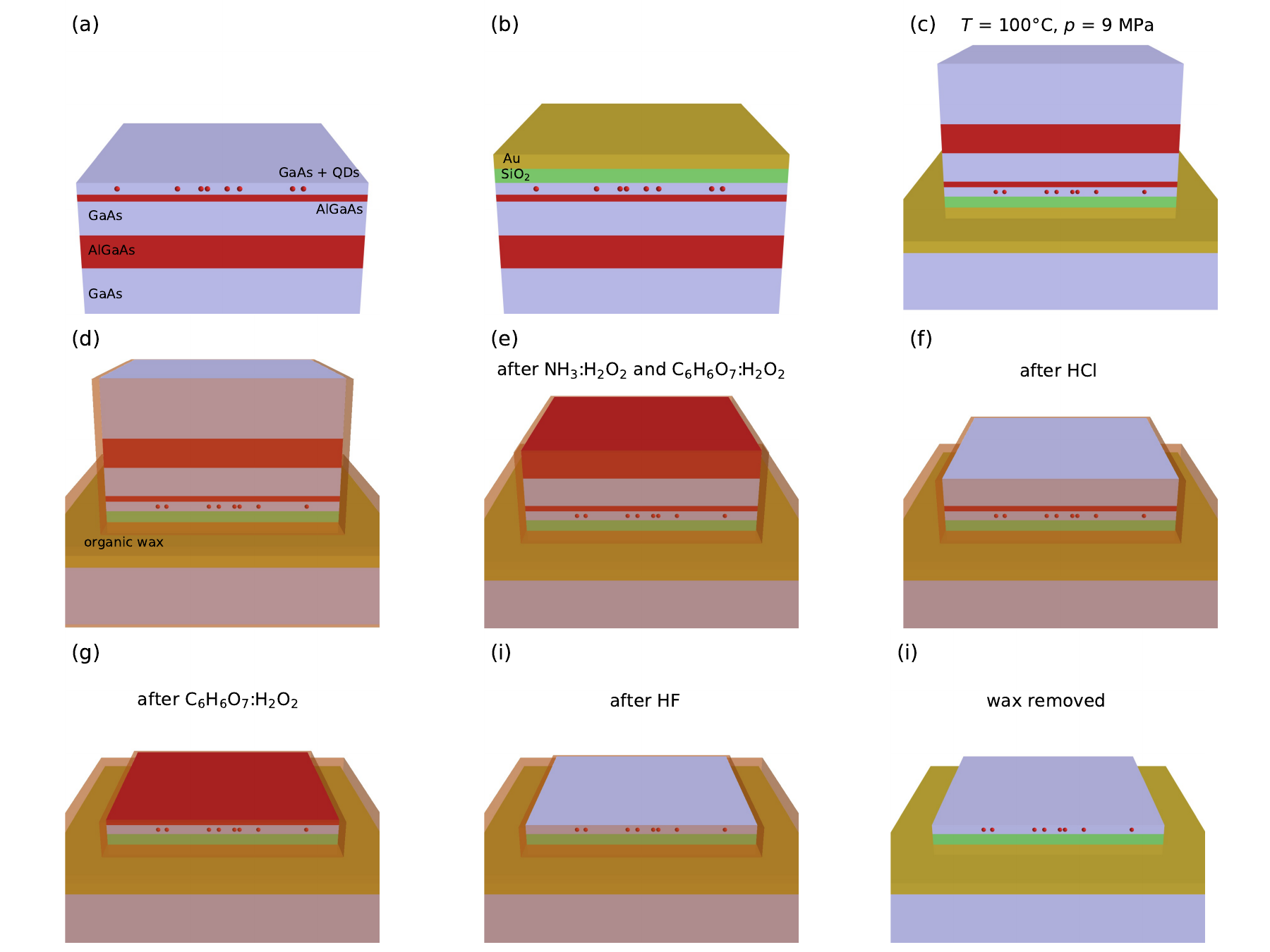}
	\caption{Schematic representation of the flip-chip process: (a) As grown wafer. (b) After dielectric and gold deposition. (c) After Au-Au thermocompression bonding. (d) Applying protective wax. (c) After ammonia-peroxide and citric acid/peroxide etching. (f) after hydrochloric acid etching. (g) After another citric-acid/perodixde etching. (i) after hydrofluoric acid etching. (i) flip-chip substrate after removal of protective wax.}
\label{fig:figureS2}
\end{figure*}

\noindent A schematic depiction of the used flip-chip substrates to fabricate the QD-hCBG cavities is shown in Figure~\ref{fig:figureS2}. The initial substrate (a) is grown on a (100)-GaAs-wafer substrate using a molecular beam epitaxy system. The growth consists of a Al$_{0.92}$Ga$_{0.08}$As layer of 500~nm thickness (first etch stop layer), followed by 500~nm of GaAs. On top of that, another 100~nm thick Al$_{0.92}$Ga$_0.08$As (second etch stop layer) is grown. The future hCBG membrane consists of a 170~nm thick GaAs layer with Stranski-Krastanov grown InAs QDs situated at its center. (b) On top of this flip-chip substrate, 200~nm of SiO$_2$ are deposited using PECVD (plasma-enhanced chemical vapour deposition) and 150~nm of gold by electron-beam evaporation. (c) The SiO$_2$-Au covered substrate is then flipped and thermo-compression bonded to a gold-coated counter-substrate at 100°C at a pressure of 9~MPa for about 2~hrs. (d) Organic wax is applied to cover all exposed areas of the bonded substrate, except for the top. (e) The underlying GaAs wafer is etched using a two-step etching solution of ammonia/hydrogen-peroxide and citric-acid/hydrogen-peroxide. (f) The first etch stop layer is removed with hydrochloric acid. (g) The intermediate GaAs layer is removed using citric-acid/hydrogen-peroxide. (i) The second etch stop layer is removed with hydrofluoric acid. (i) The final flip-chip substrate is obtained after removing the protective wax.

\clearpage
\section*{S3: Deterministic fabrication using marker based CL/EBL}
\label{sec:si_fab}

\begin{figure*}[ht]
    \center
	\includegraphics[width=0.95\textwidth]{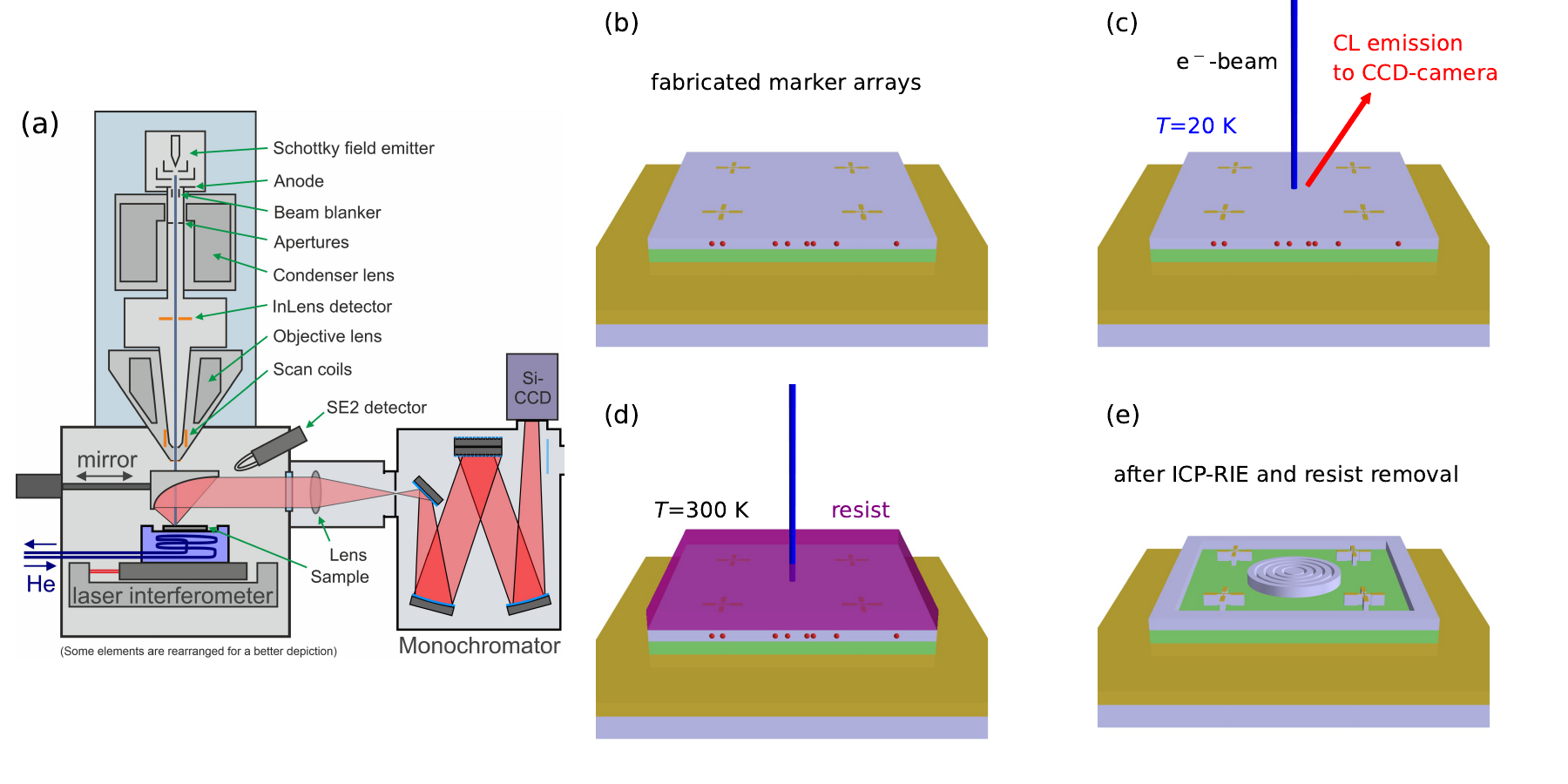}
	\caption{Overview of the deterministic fabrication process: (a) Schematic depiction of the used CL/EBL system for the deterministic fabrication of QD-hCBG cavities. (b) Marker arrays are fabricated on suitable flip-chip substrates with a lift-off process. (c) Marker arrays are scanned at $T$=4~K using the CL/EBL system in (a), while the emitted CL signal is recorded on the CCD-camera. (d) At room temperature, the mapped flip-chip substrate is spin-coated with a e-beam resist and an electron beam lithography process is carried out on the pre-determined marker arrays, with positions according to the maps recorded in (c). (e) The exposed substrate is treated with a suitable developer, and the written structures are transferred to the underlying CBG membrane with a ICP-RIE (inductively coupled plasma reactive ion etching) process. Afterwards, the residual resist is removed.}
\label{fig:figureS3}
\end{figure*}

\noindent Figure~\ref{fig:figureS3} gives a summary of the used deterministic fabrication process. The used CL/EBL system is shown in Fig.~\ref{fig:figureS3}(a). Marker arrays are fabricated on the QD-containing flip-chip substrate (Fig.~\ref{fig:figureS3}(b)), and it is placed in the CL/EBL system that allows cooling of the substrates to 20~K, and recording of CL spectra with an added spectrometer and Si-CCD.  The electron beam is scanned over the sample, and in addition to SEM images, the CL signal emitted by the sample is recorded via a 1200g/mm grating on a CCD-camera (Fig.~\ref{fig:figureS3}(c)). The QD's position is extracted via a 2D-Gaussian fit, and is determined relative to the centers of the four markers for each investigated marker arrays. The central marker position can be determined with $<$10~nm precision. Fitting accuracy of the QD from its 2D-Gaussian profile can be as good as $<$5~nm, depending on the QD's properties. The largest impact on the QD's position has the density of emitters on the sample, both spatially and spectrally.     

The sample is then warmed up and taken out of the system, in order to spin-coat it with an electron-sensitive resist and perform an EBL process to write the hCBG cavities based on the previously determined QD positions per marker-array (Fig.~\ref{fig:figureS3}(d)). This EBL is performed with the same system used for the CL-mapping, ensuring that the marker recognition depends on the same resolution as for the QD identification. The exposed sample is then developed and inductively-coupled plasma reactive ion etching (ICP-RIE) is used to transfer the CBG pattern written in the resist into the GaAs membrane, while the SiO$_2$ layer serves as a underlying etch-stop layer. The residual resist is removed in a final step (Fig.~\ref{fig:figureS3}(d)). 

\clearpage
\section*{S4: QD integration accuracy}
\label{sec:si_integration}

\setcounter{figure}{0}
\renewcommand{\figurename}{Fig.}
\renewcommand{\thefigure}{S4-\arabic{figure}}

\begin{figure*}[ht]
    \center
	\includegraphics[width=1\textwidth]{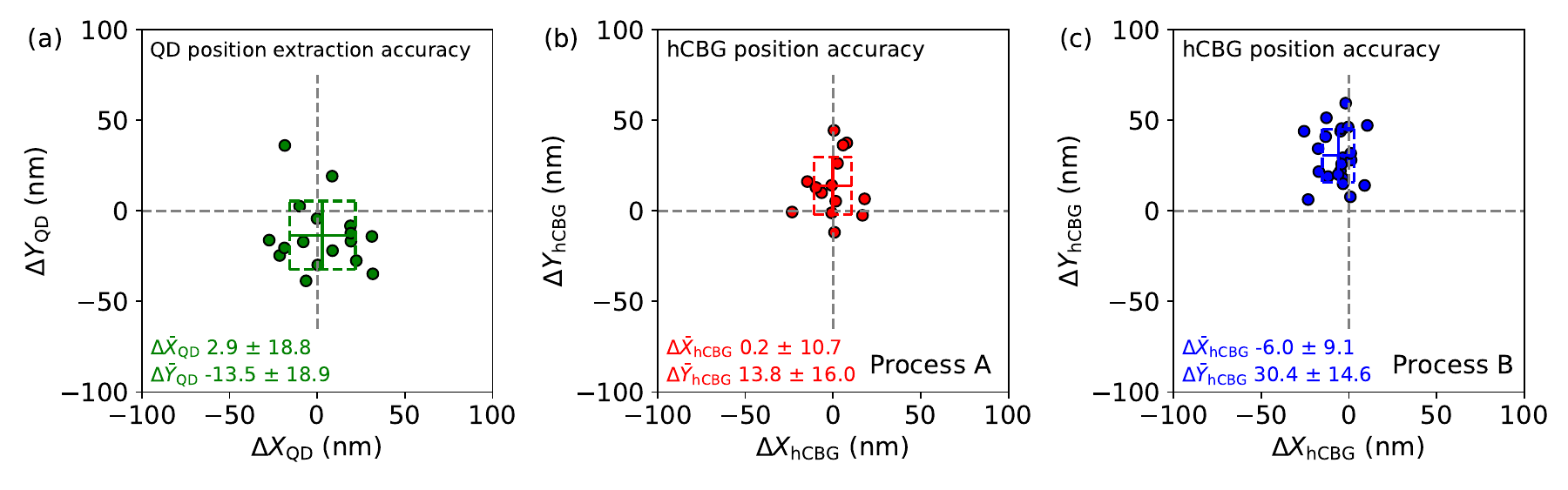}
	\caption{Position deviations $\mathrm{\Delta}X$ and $\mathrm{\Delta}Y$ in $x-$ and $y-$ direction between two respective consecutive CL scans on several marker arrays for (a) a planar hCBG substrate to estimate the QD position extraction accuracy, and (b, c) the deviation between determined QD position before and extracted hCBG cavity position after the deterministic integration.}
\label{fig:figureS3b}
\end{figure*}

\noindent The overall spatial integration accuracy of a target QD emitter into the fabricated hCBG cavity $\sigma_\mathrm{spatial}$ consists of the accuracy to determine the QD's position $\sigma_\mathrm{QD}$ and the accuracy to fabricate the hCBG cavity at the target position $\sigma_\mathrm{hCBG}$:
\begin{equation}
    \sigma_\mathrm{spatial} = \sqrt{\sigma_\mathrm{QD}^2+\sigma_\mathrm{hCBG}^2}.
    \label{eq:sigma_spatial}
\end{equation}
We estimate the accuracy of extract the QD's position form the CL-maps by recording the QD-emission in CL maps from the same marker array two consecutive times and compare the extracted QD-position from their emission. Fig.~\ref{fig:figureS3b}(a) shows the deviations $\mathrm{\Delta}X_\mathrm{QD}$ and $\mathrm{\Delta}Y_\mathrm{QD}$ in $x$- and $y$-direction between these consecutive maps, and the respective average values with standard deviation. We obtain $\sigma_\mathrm{QD}$ from the standard deviations as $\sigma_\mathrm{QD}=\sqrt{18.8^2+18.9^2}$ = 26.7\,nm. 

\begin{figure*}[ht]
    \center
	\includegraphics[width=1\textwidth]{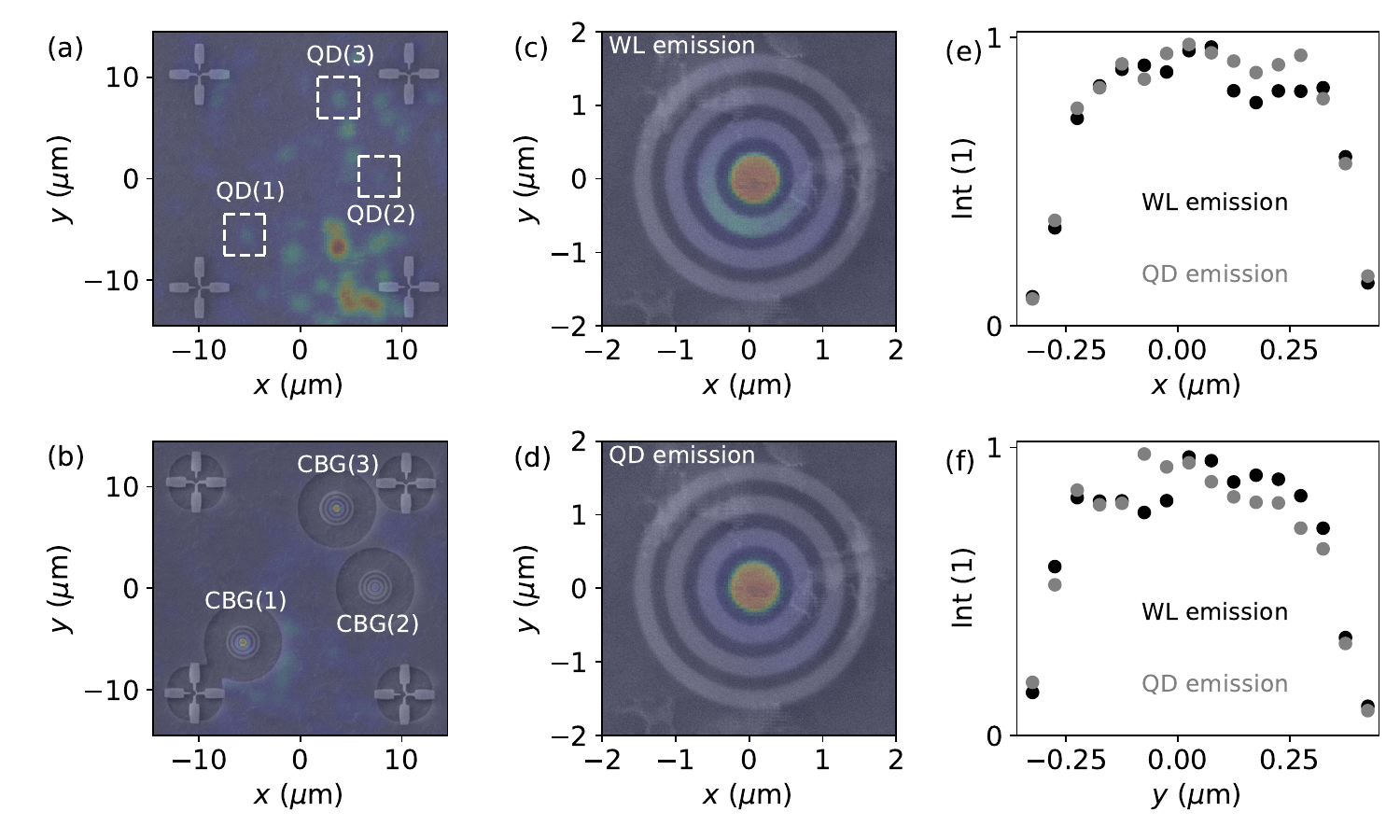}
	\caption{Measurements to determine the hCBG positioning accuracy. (a) CL map recorded before and (b) after the integration of selected QDs (1-3) into CBG cavities(1-3). (c) Close-up CL scan of an hCBG cavity with integrated wetting layer emission and (d) target QD emission visible as heatmap. (e) Corresponding integrated intensity across $x$- and (f) $y$-direction of the central disc for both wetting layer and QD emission. The respective mean deviations and standard deviations in $x$- and $y$- are indicated, and the accuracies are estimated from the standard deviations (see text).}
\label{fig:figureS3c}
\end{figure*}

To gain information on $\sigma_\mathrm{hCBG}$, we record CL maps on marker-arrays with deterministically fabricated QD-hCBG cavities (see Figure~\ref{fig:figureS3c}(a) and (b)). Due to the small size of the hCBG's central mesa, the QD's emission is much more confined to than for the planar maps. The spatial emission distribution in this case does not correspond to the QD itself anymore, but it rather "fills" the entire hCBG's central mesa. This is apparent since the spatial distribution of the emission stays the same, regardless if the QD's spectral region is integrated, or if the wetting layer emission is integrated (see Figure~\ref{fig:figureS3c}(c-f)). Fitting the spatial emission distribution for the CL-maps after the integration therefore yield an position of the hCBG's central mesa, and can thus be used to determine how accurately the hCBG cavity can be placed at a target location.

Figures~\ref{fig:figureS3b}(b) and (c) show the deviations $\mathrm{\Delta}X_\mathrm{hCBG}$ and $\mathrm{\Delta}Y_\mathrm{hCBG}$ in $x$- and $y$-direction for the fabricated hCBG position of two Processes A and B using this method, compared to the intended position. The yield $\sigma_\mathrm{hCBG,A} = \sqrt{10.7^2+16.0^2}$ = 19.2\,nm and $\sigma_\mathrm{hCBG,B} = \sqrt{9.1^2+14.6^2}$ = 17.2\,nm, yielding an average accuracy of the hCBG cavity positioning of $\sigma_\mathrm{hCBG} = 18.2$\,nm.
\\

\noindent With the two obtained values for QD position determination accuracy and hCBG cavity positioning accuracy according to eq.~\ref{eq:sigma_spatial}), we estimate an overall integration accuracy of $\sigma_\mathrm{spatial}=\sqrt{26.7^2+18.2^2} = 32.3$\,nm. This is the value indicated in Fig.~1(f) in the main manuscript. 

We note that the QD position determination accuracy and hCBG positioning accuracy for the reference processes shown in Fig.~\ref{fig:figureS3b} appear to have a larger spread in $y$-direction. We aim to investigate and reduce this in the future and currently suspect a combination of residual drift and charging effects. We believe the influences of both of these factors to be more pronounced in y-direction, since the CL-mapscans are currently performed as $x$-linescans for subsequent $y$-pixels, meaning that neighbouring $y$-pixels are scanned at longer times apart.
\clearpage

\section*{S5: Micro-photoluminescence setup}
\label{sec:si_setup}

\setcounter{figure}{4}
\renewcommand{\figurename}{Fig.}
\renewcommand{\thefigure}{S\arabic{figure}}

\begin{figure*}[ht]
    \center
	\includegraphics[width=0.6\textwidth]{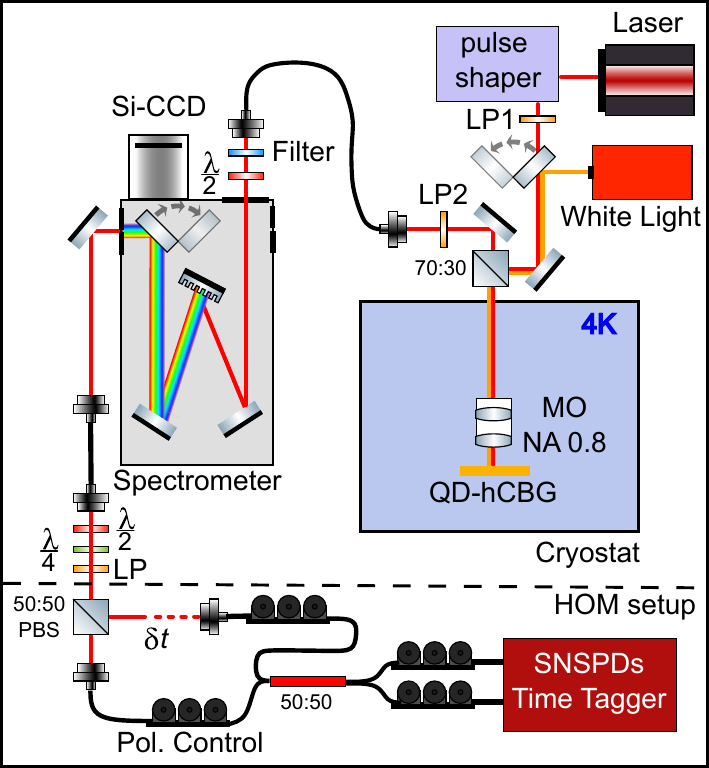}
	\caption{Overview of the used $\mu$-PL setup. MO: microscope objective, LP: linear polarizer, $\frac{\lambda}{2}$: quarter waveplate, $\frac{\lambda}{4}$: quarter waveplate.}
\label{fig:figureS4}
\end{figure*}

\noindent The used optical $\mu$-PL setup for all measurements is depicted schematically in Figure~\ref{fig:figureS4}. The QD-hCBG cavity sample is placed in a closed-cycle helium cryostat and cooled to 4~K with a low-temperature compatible microscope objective with NA=0.8 for confocal excitation/emission experiments. Excitation light reaching the sample via the excitation path is either from a pulsed 80~MHz laser system with nominal 2~ps temporal pulse-width (picoEmerald, APE GmbH) or from a broadband white-light source (SLS201L/M, Thorlabs GmbH). The pulsed laser passes through a commercial pulse shaper to change the pulse's spectral and temporal width. For the experiments under strictly resonant excitation, a linear polarizer is placed in the excitation and detection path, respectively, to enable measurements in cross-polarization. The emission from the sample is coupled into a single mode fiber (780HP) path, and coupled-out towards a 0.75~m spectrograph (entrance slit width: 100~$\mu$m) with 1200~g/mm grating, using a halfwaveplate before the monochromator to match the polarization to the grating. For p-shell experiments, the emission from the sample is filtered with a bandpass filter ($\lambda_\mathrm{center}=940 \pm 10$~nm) before entering the monochromator. To spectrally resolve the signal, the emission is reflected from the grating onto a 1340~pixel Si-CCD camera cooled to 200$\,$K with a Peltier element. For time-resolved measurements, the emission is reflected from the grating trough a separate spectrograph's exit (exit slit width: 80~$\mu$m) into a single mode fiber (SMF 780HP).
The SMF leads to a Hong-Ou-Mandel setup set as follows: An initial free-space polarization control consisting of half-waveplate, quarter-waveplate and linear polarizer ensures equal splitting at a polarizing BS to match both arms of the Mach-Zehnder interferometer (MZI) in intensity. The arms of the MZI are precisely time-matched for the time-difference $\delta{t}$ ($\delta{t}=12.5$~ns for 80~MHz excitation rate, and $\delta{t}=781$~ps for 1.28~GHz excitation rate). The interference then takes place in fiber on a 50:50 fiber-beamsplitter, where the input polarizations are fully-controlled by polarization paddles in fiber to set the co- and cross-polarized interference. The output of the 50:50 beam-splitter leads to superconducting nanowire single photon detectors (SNSPDs) (Single Quantum EOS CS, Single Quantum B.V.) connected to time tagging electronics (QuTag, qtools GmbH). Additional polarization paddles on the fibers leading to the detector match the polarization to the nanowire orientation. The correct setting of polarization and time delay in the MZI is confirmed by interfering laser pulses at the QD emission wavelength and observing close to 100$\%$ interference contrast.

\clearpage

\section*{S6: QD state identification}
\label{sec:si_qdstates}

\begin{figure*}[ht]
    \center
	\includegraphics[width=1\textwidth]{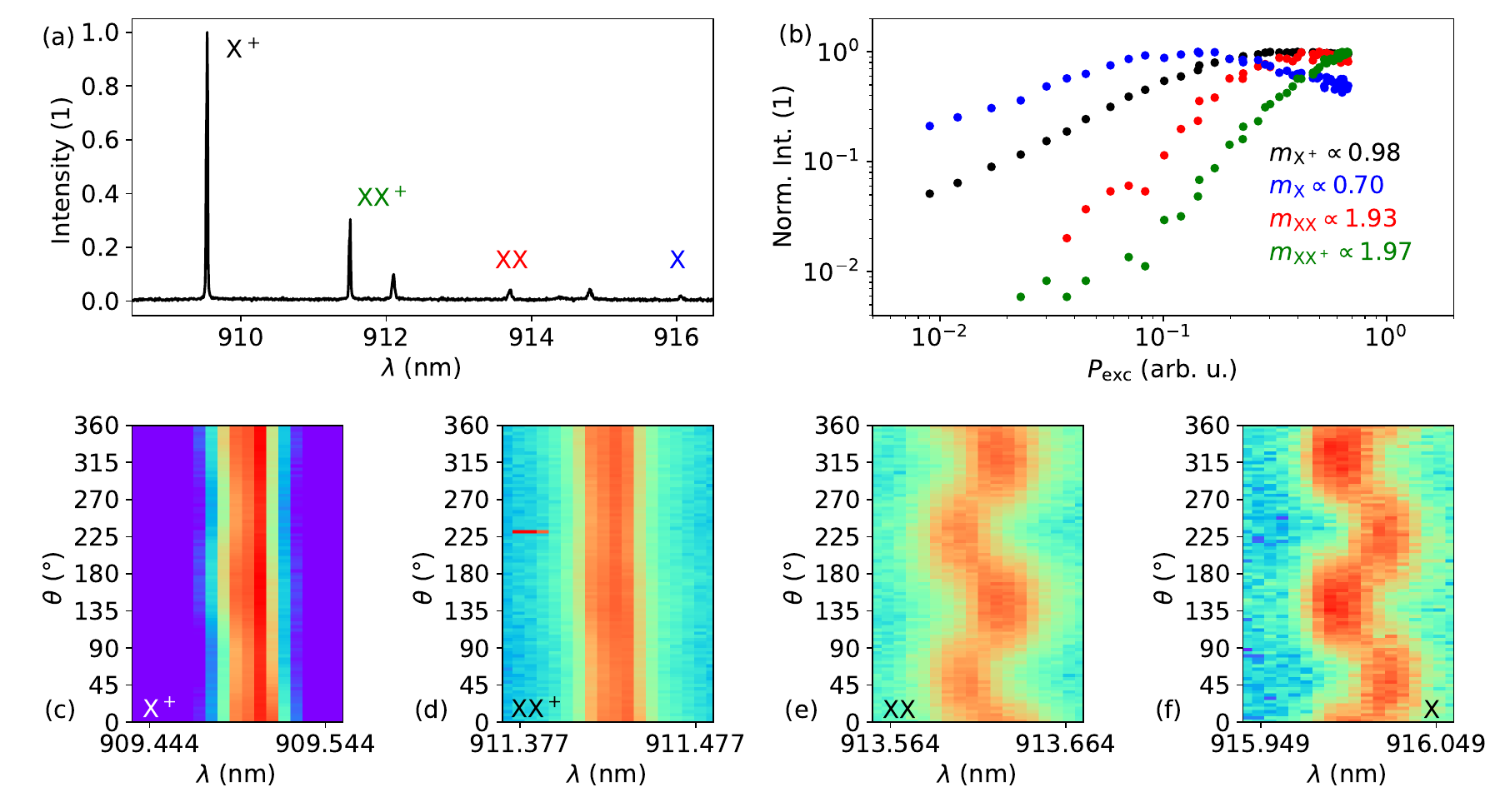}
	\caption{\textbf{Excitation power- and polarization-dependent measurements to identify QD states.} (a) PL-spectrum of exemplary QD with emission lines assigned to indicated QD states. (b) Excitation power-dependent PL-intensities of the indicated emission lines, with extracted slopes $m$. (c)-(f) Polarization-dependent PL-intensities for the indicated emission lines.}
\label{fig:figureS5a}
\end{figure*}

\noindent Figure~\ref{fig:figureS5a}(a) shows a typical spectrum of the studied InAs/GaAs QDs. The charged states X$^+$ and XX$^+$ as indicated in Fig.~\ref{fig:figureS5a}(a) are distinguished from the neutral X and XX states by polarization resolved measurements shown in Fig.~\ref{fig:figureS5a}(c)-(f): For X and XX, the fine-structure-splitting (FSS) is clearly apparent from the polarization series, while X$^+$ and XX$^+$ show no resolvable FSS, indicating that they are charged states of the QD. The respective one-particle states X and X$^+$ are distinguished from XX and XX$^+$ based on their lower slope in the power-dependent measurements shown in Fig.~\ref{fig:figureS5a}(b). Note that XX$^+$ exhibits two optically active transitions and only the one at $\sim911.4$\,nm is shown in the power- and polarization dependent measurements here. X$^+$ is identified as a positive charged state due to the XX$^+$ appearing blue-shifted from the neutral exciton, indicating that positively charged states in the sample exhibit anti-binding characteristics, as commonly observed in InAs QDs (see. e.g. Rodt \textit{et al.}, Phys. Rev. B. 71, 155325 (2005)).

\clearpage
\section*{S7: PLE and quasi-resonant excitation properties of planar QDs}
\label{sec:si_PLE}

\begin{figure*}[ht]
    \center
	\includegraphics[width=0.65\textwidth]{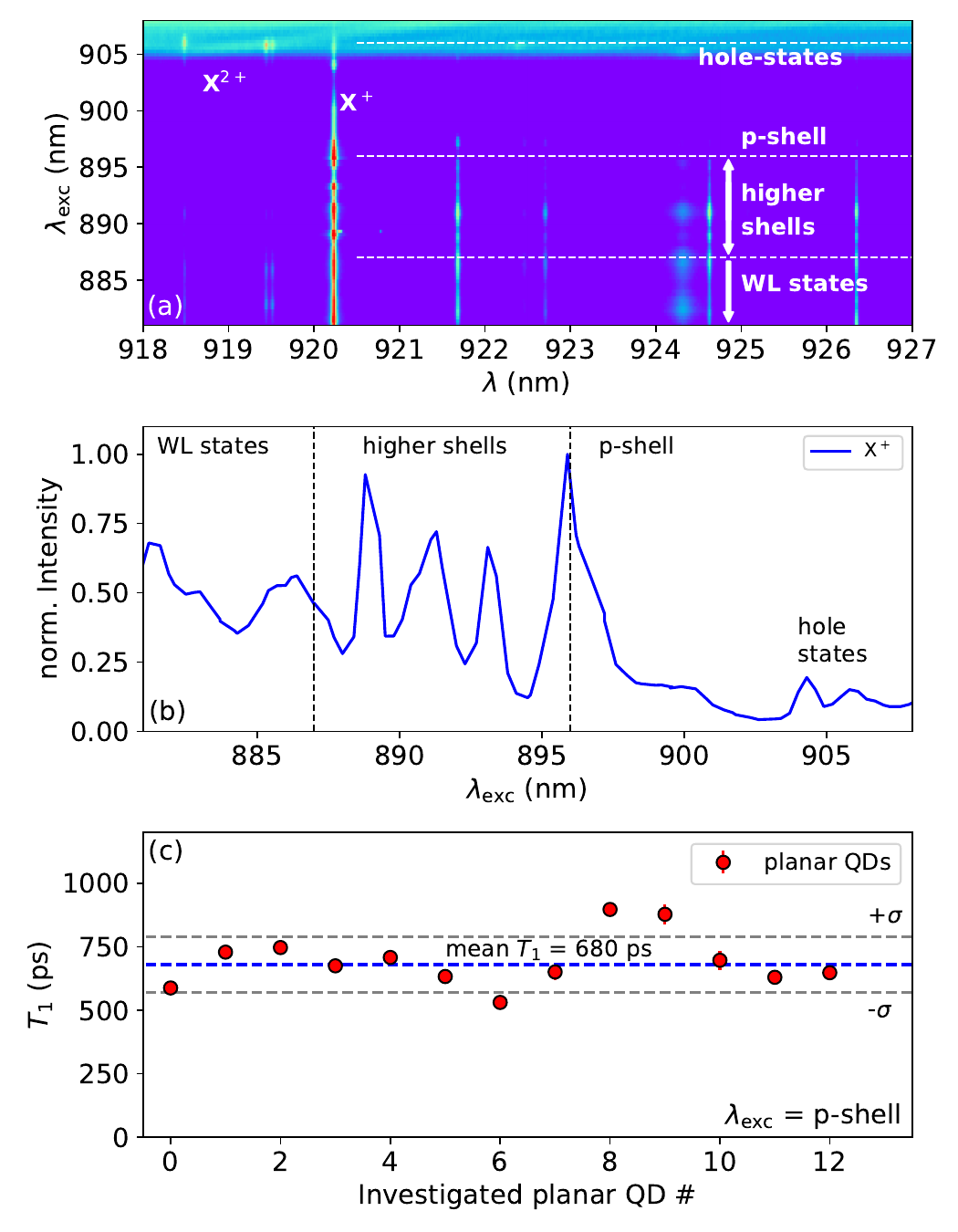}
	\caption{Additional data on quasi-resonant excitation. (a) PLE (intensity in logarithmic scale) of the QD-hCBG cavity in Fig.~\ref{fig:figure3} in the main text. Indicated is a typical p-shell for the investigated InAs QDs. The PLE shows several quasi-resonances, which can be attributed to QD shells, excitation in the wetting layer or different interactions with hole-states. The increased intensity at $\lambda_\mathrm{exc}>905$~nm originates from insufficient laser suppression. (b) PLE resolved for the X$^+$-state from (a). (c) $T_1$-times of the X$^+$ states in numerous QDs from the surrounding membrane area (i.e., not integrated in a hCBG-cavity) under p-shell excitation. Indicated are the positive and negative standard-deviation $\sigma$. The corresponding mean $T_1$ for a planar dot is 680~ps.}
\label{fig:figureS5}
\end{figure*}

\clearpage
\section*{S8: Rabi rotations under RF excitation}
\label{sec:si_RF}

\begin{figure*}[ht]
    \center
	\includegraphics[width=0.65\textwidth]{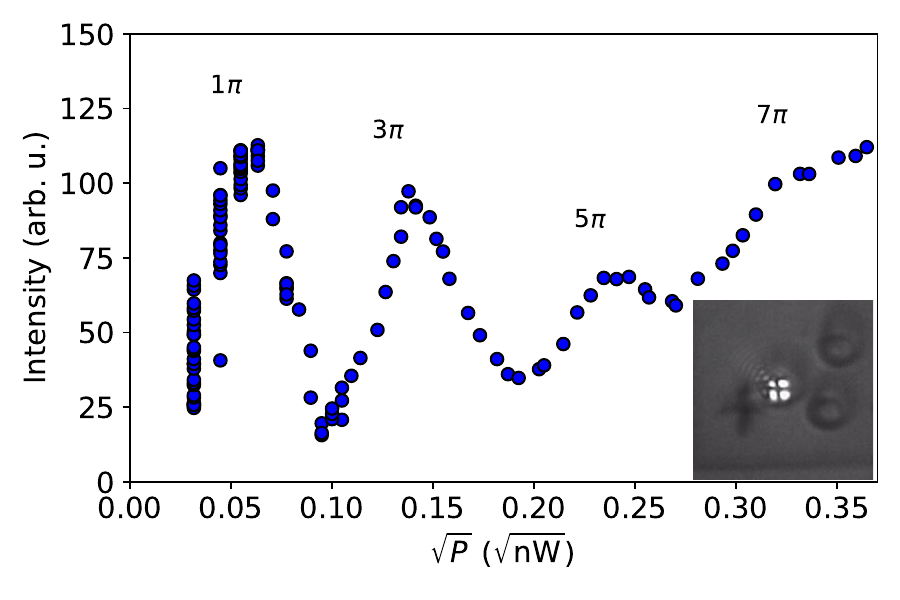}
	\caption{Power dependent X$^+$ intensity under s-shell excitation, showing Rabi rotations corresponding to subsequent pi-pulses. The inset shows the "clover leaf" pattern of the excitation spot on a QD-hCBG cavity when excited in cross-polarized excitation. }
\label{fig:figureS6}
\end{figure*}

\clearpage
\section*{S9: Temperature dependent TPI}

\begin{figure*}[ht]
    \center
	\includegraphics[width=1\textwidth]{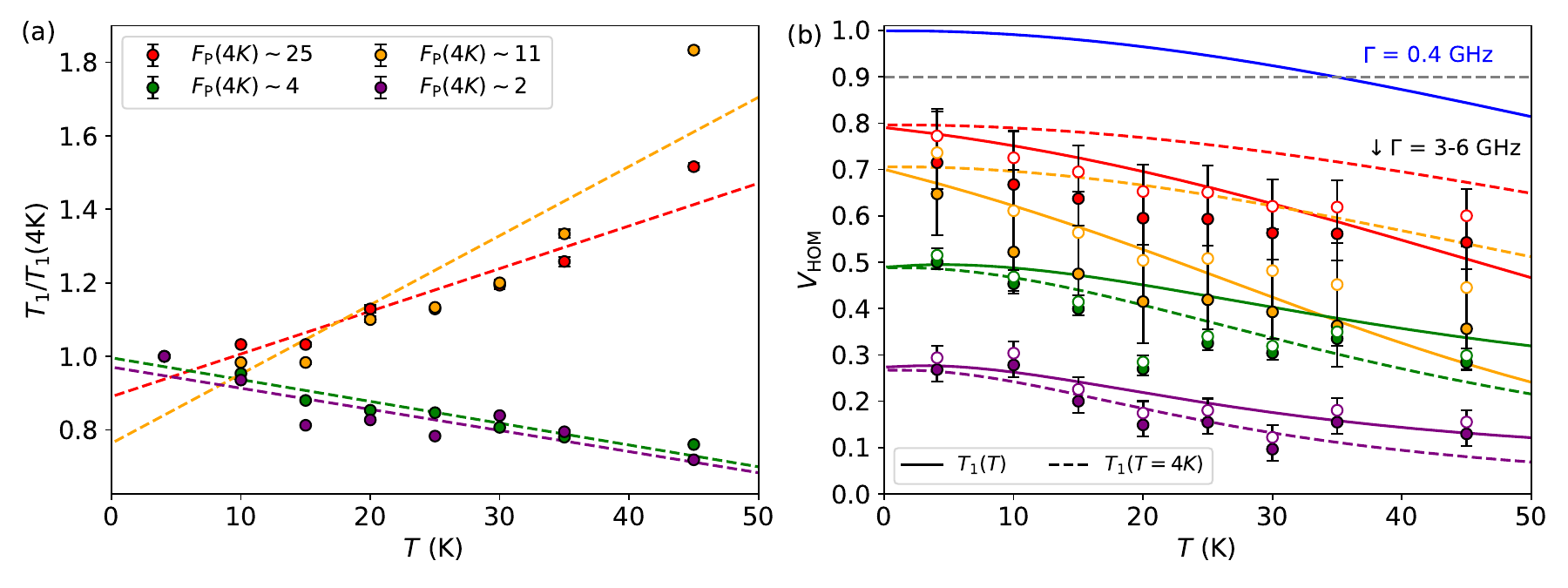}
	\caption{Further details on the temperature dependent $V_\mathrm{HOM}$-measurements shown in Fig.~4(d) in the main manuscript. (a) Extracted $T_1$-times for the four investigated cavities for varying temperatures The dashed lines correspond to linear fits. (b) Data and theory curves (solid lines) displayed in Fig.~4(d) in the main manuscript. The dashed lines correspond to predicted $V_\mathrm{HOM}(T)$ if the $T_1$-time could be held kept at its value at $T=5$\,K for each of the cavities. The blue line depicts theoretical $V_\mathrm{HOM}(T)$ for $F_\mathrm{P}=25$ and $\Gamma=400$\,MHz, as can be expected from coherent s-shell excitation.}
\label{fig:figureS9}
\end{figure*}

\noindent As pointed out in the main manuscript, the temperature dependent $V_\mathrm{HOM}$-measurements were impacted by a $T$-dependent $T_1$-change due to the spectral shift of the respective X$^+$-lines relative to the cavity modes. The resulting varying $T_1$-times exctracted from time-resolved measurements are shown in Fig.~\ref{fig:figureS9}(a). The $T_1(T)$-dependency is approximated with linear fits shown as dashed lines, and is used in eq.~(2) in the main manuscript to calculate $V_\mathrm{HOM}(T_1,\Gamma)$ and $\gamma(T_1)$ for the given $T$-range. The resulting theoretical $V_\mathrm{HOM}(T)$ is shown with the experimental data in Fig~4(d) in the main manuscript, and also in Fig.~\ref{fig:figureS9}(b) together with the experimental (corrected) data. 

The dashed lines correspond to theoretical $V_\mathrm{HOM}(T)$ if the $T_1$-time obtained at $T=5$\,K for each of the cavities could be maintained for the whole investigated $T$-range. Comparison of the dashed and solid lines for the four cavities show, that the high $F_\mathrm{P}$=25 (red) and 11 (orange) cavities were primarily limited by the $T_1(T)$-change over temperature for their $V_\mathrm{T}$-values, and would benefit greatly for an independent tuning mechanism that would allow to control $T_1$ indepedently from temperature. Likewise, the lower $F_\mathrm{P}$=4 (green) and 2 (purple) cavities benefited from an increased Purcell enhancement due to better spectral matching of their X$^+$ to the cavity mode at elevated temperature, masking the influence of the increase phonon interactions on their measured $V_\mathrm{HOM}(T)$.

The blue line shown in Fig.~\ref{fig:figureS9}(b) shows predicted $V_\mathrm{HOM}(T)$ according to eq.~(2) in the main manuscript for $F_\mathrm{P}$=25 and an inhomogenous broadening of $\Gamma=400$\,MHz, as was extracted from the s-shell excited TPI measurements in the main manuscript. It is apparent that for excitation methodes with higher coherence than the quasi-resonant p-shell excitation employed in the temperature-dependent TPI measurements, high $F_\mathrm{P}$ QD-hCBG devices promise $V_\mathrm{HOM}>0.9$ at temperatures as high as 30\,K.

\clearpage
\section*{S10: Rate multiplication setup for GHz excitation}
\label{sec:si_GHz}

\begin{figure*}[ht]
    \center
	\includegraphics[width=0.55\textwidth]{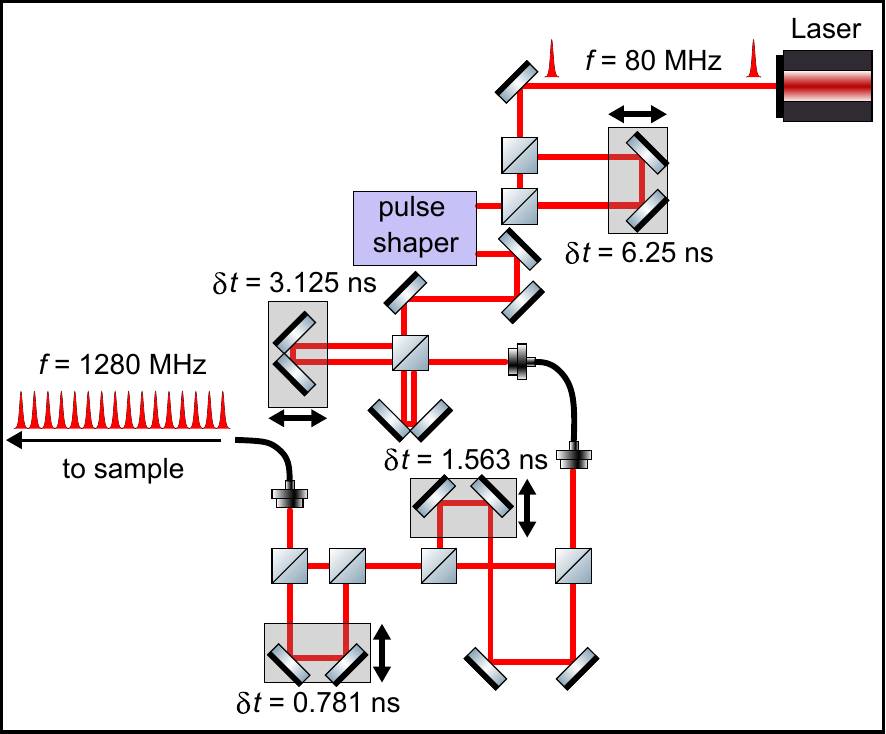}
	\caption{Excitation rate multiplication setup used for GHz excitation rates from a pulsed 80~MHz laser by subsequent beamsplitters and delay stages. Every successive delay stage is time-controlled via movable precision stages, and every delay stage doubles the previous frequency. The powers of the multiplied pulses can be precisely controlled with the alignment of the delay paths and additional filters. Due to the splitting of the initial input power, the maximum reachable power per pulse in p-shell excitation is $\sim1$~$\mu$W at the sample, far from saturation of the QD at these conditions. In s-shell excitation, this power is more than enough to achieve $\pi$-pulse conditions.}
\label{fig:figureS7}
\end{figure*}

\end{document}